\documentclass[graybox, envcountchap]{svmult}

\newcommand{\apj}{Astrophys.\ J.\ }
\newcommand{\apjl}{Astrophys.\ J.\ }
\newcommand{\apjs}{Astrophys.\ J. Suppl.\ }
\newcommand{\aap}{Astron.\ Astrophys.\ }

\newcommand{\araa}{Annu.\ Rev.\ Astron. Astrophys.\ }

\newcommand{\nat}{Nature~}

\newcommand{\mnras}{Mon.\ Not.\ R.\ Astron.\  Soc.\ }
\newcommand{\pasj}{Publ.\ Astron.\ Soc.\ Japan\ }

\newcommand{\aapr}{Astron.\ Astrophys.\ Rev.\ }

\newcommand{\na}{New Astron.\ }
\newcommand{\nar}{New Astron.\ Rev.\ }

\usepackage{mathptmx}        
\usepackage{amsmath}
\usepackage{amssymb}
\usepackage{color}
\usepackage{helvet}          
\usepackage{courier}         
\usepackage{dirtree}

\usepackage{makeidx}        
\usepackage{graphicx}        
\usepackage{subfig}

\usepackage{multicol}        
\usepackage[bottom]{footmisc}

\usepackage{hyperref}        
\hypersetup{colorlinks=true,urlcolor=blue}
\usepackage{cite}

\makeindex             

\begin{document}


\title{Tilted Accretion Disks}
\author{P. Chris Fragile and Matthew Liska}
\institute{P. Chris Fragile \at College of Charleston, 66 George Street, Charleston, SC, USA, \email{fragilep@cofc.edu}
\and Matthew Liska \at Georgia Institute of Technology, Howey Physics Bldg, 837 State St NW, Atlanta, GA, USA, \email{mliska3@gatech.edu}}
%
%
\maketitle

\abstract{In this chapter, we review some of the interesting consequences that tilt between the spin axis of the black hole and angular momentum axis of the accretion disk can have on the dynamics, thermodynamics, and observational appearance of accreting systems, from precessing coronae and jets to standing nozzle shocks and quasi-periodic oscillations. We begin the chapter by examining some of the reasons tilted disks are interesting as well as present arguments for how ubiquitous they may be. We then review the existing simulation results in the literature, broadly dividing them into sections on thick disks, thin disks, and magnetically arrested disks (MADs). We finish by highlighting some of the phenomenology that is unique to tilted disk simulations and discuss how this may apply to observations.}


\section{General Overview}
\label{sec:overview}

\subsection{Why is Tilt Interesting?}

\begin{figure}
\sidecaption
\includegraphics[width=1.0\linewidth,trim=0mm 0mm 0mm 0,clip]{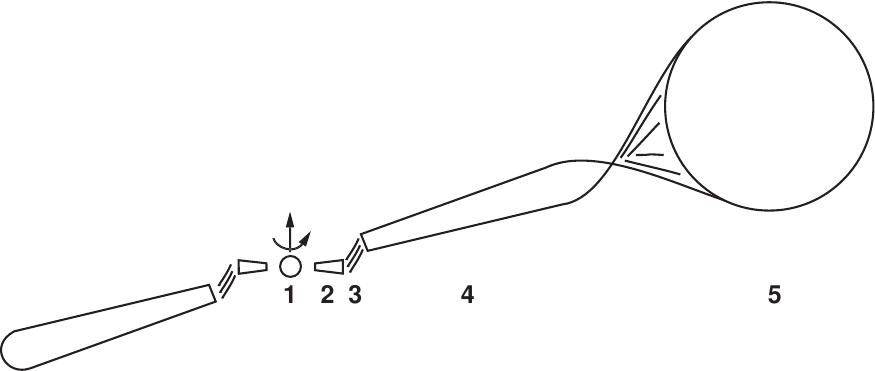}
%
%
\caption{Schematic diagram of the Bardeen-Petterson effect in a black hole X-ray binary showing (1) the central rotating black hole, (2) the inner, aligned accretion disk, (3) the transition region, (4) the outer, tilted accretion disk, and (5) the companion star. Image reproduced by permission from \cite{Fragile01}, copyright by AAS.}
\label{fig:BP}       
\end{figure}

If an accretion disk is misaligned, or ``tilted,'' with respect to a compact rotating object, it will be subject to Lense-Thirring (LT) precession \cite{Lense18}. For an ideal test particle in a slightly tilted orbit at a radius $r$ around a black hole of mass $M$ and specific angular momentum $a_* = J/M^2 = a/M$, this precession occurs at an angular frequency $\Omega_\mathrm{LT} \approx 2a_*M/r^3$ (in units where $G=c=1$). Close to the black hole, this is comparable to the orbital angular frequency $\Omega = (M/r^3)^{1/2}[1+a_*(M/r^3)^{1/2}]$. However, because of its strong radial dependence, LT precession becomes much weaker far from the hole. Therefore, a tilted disk made of non-interacting rings would experience differential precession that would twist and warp it over time, or even tear it if the misalignment was large enough \cite{Nixon12b}. However, in a real disk, there can be strong interactions between successive radial rings, such that any warping disturbance will be communicated over some range of the disk. This communication is generally thought of as happening in either a diffusive or wave-like manner, depending on the properties of the disk. In the diffusive case, the warping is limited by secular (i.e., ``viscous'') responses within the disk. In such a case, the misaligned angular momentum is transported outward by viscous torques until the timescale for outward transport roughly equals the precession timescale \cite{Kumar85, Nelson00}, leading to a unique, nearly constant transition radius $r_\mathrm{BP}$ \cite{Bardeen75,Kumar85}, inside of which the disk is expected to be flat and aligned with the black-hole symmetry plane, and outside of which the disk is also expected to be flat but in a plane determined by the angular momentum vector of the larger gas supply. This is often termed the ``Bardeen-Petterson'' (BP) effect and is illustrated in Figure \ref{fig:BP}. 

The BP effect is expected to apply whenever the dimensionless stress parameter $\alpha$ within the disk is larger than the scale height $\delta = H(r)/r$, where $H(r)$ is the thermal semi-thickness of the disk. Given that $\alpha$ is usually thought to be significantly less than one, this implies geometrically thin disks. However, neglecting relativistic correction factors, the scale height in the innermost, radiation-pressure- and electron-scattering-dominated portion of thin disks is roughly given by
\[
\delta \sim \eta^{-1} \left(\frac{L}{L_\mathrm{Edd}}\right)\left(\frac{r}{r_G}\right)^{-1} ~,
\]
where $\eta \sim 0.1$ is the radiative efficiency, $L/L_\mathrm{Edd}$ is the luminosity in units of Eddington, and $r_G = GM/c^2$ is the gravitational radius. This means that the BP effect will only be relevant for the inner regions of thin disks at very small Eddington ratios, $L/L_\mathrm{Edd} \le \alpha \eta \ll 1$, but at those luminosities, real black hole accretion systems are generally found in the hard state, so the thin-disk solution may not be applicable. Furthermore, other accretion solutions, such as radiatively inefficient, geometrically slim and thick flows, will clearly not be in the Bardeen-Petterson regime. Therefore, the prevalence of this configuration may be limited. Nevertheless, we review numerical results applicable to the Bardeen-Petterson limit in Section \ref{sec:BPeffect}.

\begin{figure}
\sidecaption
\includegraphics[width=1.0\linewidth,trim=0mm 0mm 0mm 0,clip]{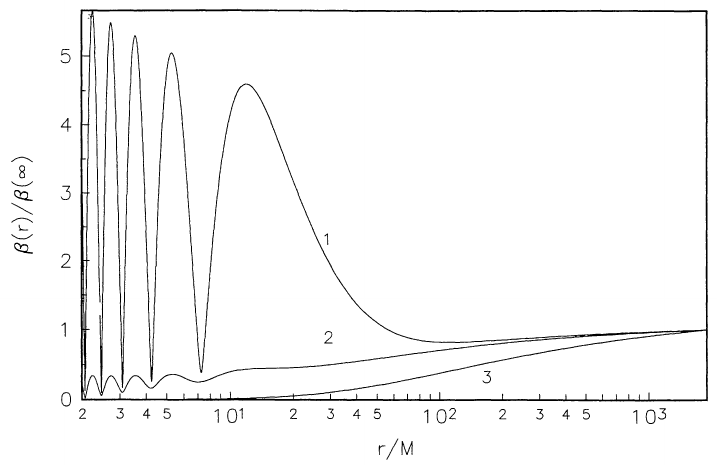}
%
%
\caption{Radial dependence of the tilt angle $\beta(r)/\beta(\infty)$ for values of $\delta/\alpha$ of 10 (curve 1), 1 (curve 2), and 0.1 (curve 3). Curve 1 would be in the wave-like limit, whereas curve 3 would be in the Bardeen-Petterson limit. Image reproduced by permission from \cite{Ivanov97}, copyright by RAS.}
\label{fig:Ivanov}       
\end{figure}

In the thick-disk, $\delta > \alpha$ regime, LT precession is expected to produce warps that propagate in a wave-like manner \cite{Papaloizou95}, ultimately developing a standing, bending-wave pattern \cite{Ivanov97}, where the tilt varies as a function of radius and can even take on values greater than in the outer disk, as illustrated in Figure \ref{fig:Ivanov} (curve 1). Additionally, in the thick-disk limit, when the sound-crossing time is short compared to the precession timescale, a finite, isolated disk can precess as a solid body \cite{Fragile05,Fragile07,Motta18}, which can have numerous interesting observational consequences. We discuss simulations in this regime and some of the corresponding phenomenology in Section \ref{sec:precession}.

\subsection{How Ubiquitous is Tilt?}
\label{sec:ubiquity}

Although tilt introduces new physical solutions and interesting corresponding phenomenology, there still seems to be some lingering doubt within the community as to how applicable it is to real black hole accretion systems. Hence, before proceeding with our review of tilted disk simulations, we lay out some of the arguments that support such disks being present, if not commonplace, in nature. Let's start with the simple cases: tidal disruption events (TDEs) and active galactic nuclei (AGN) after major merger events. In both cases, there is no way for the accretion flow at large scales to know about or be affected by the orientation of the spin axis of the black hole; therefore, there is absolutely no reason to expect things to be aligned at the start of such events. In fact, in such cases, retrograde accretion, where the angular momentum vector of the accreting gas points opposite the spin axis of the black hole, may be just as common as prograde \cite{King08}.

Even in stellar-mass accretion systems, misalignment may be commonplace \cite{Fragile01,Maccarone02}. Consider the formation avenues for black hole X-ray binaries (BHBs). Here, the orientation of the outer disk is fixed by the binary orbit, whereas the orientation of the black hole is set at the time it becomes part of the system. If the black hole joined the binary through multi-body interactions, such as binary capture or replacement, which may be common for X-ray binaries in globular clusters, then there would have been no preexisting symmetry, so the resulting system would nearly always harbor a tilted black hole. Even if the black hole formed from a member of a preexisting binary via a supernova, as is expected among field binaries, the black hole could still end up tilted if the supernova explosion were asymmetric \cite{Fragos10}, as it appears many are.

Following their formation, there is a tendency for tilted accretion disks to eventually align with the black hole due to their mutual tidal interactions \cite{Rees78}. However, for BHBs and TDEs, the alignment timescale is comparable to or longer than their observational lifetimes \cite{Scheuer96}, and for AGN, each merger event has the possibility of reorienting the central black hole or its fuel supply, potentially resulting in repeated tilted configurations \cite{Kinney00}. Thus, while we may not know the exact prevalence of tilted accretion disks, there are ample theoretical arguments to suggest they may be quite common.

Additionally, there is now mounting observational evidence to support the same conclusion. Several BHBs, e.g. GRO J1655-40 \cite{Orosz97}, XTE J1550-564 \cite{Hannikainen01,Orosz02}, V4641 Sgr \cite{Miller02}, GX 339-4 \cite{Miller09}, and MAXI J1820+070 \cite{Poutanen22}, and AGN, e.g. NGC 3079 \cite{Kondratko05}, NGC 1068 \cite{Caproni06}, and NGC 4258 \cite{Caproni07}, have accretion disks that appear to be tilted with respect to the symmetry plane of their central black hole spacetimes. At present, most of these claims are circumstantial, as they rely on observations of relativistic bipolar jets (assumed to be aligned with the spin axis of the black hole) that are not perpendicular to the plane of the accretion disk observed at large scales \cite{Kinney00, Schmitt02} or because the disk is warped in a way that may be consistent with the expectations of the BP effect \cite{Middleton16}. In the future, more direct evidence for tilted disks may come from polarization measurements, which can be sensitive to the inclination and viewing angle of the inner accretion disk \cite{Cheng16}.

\subsection{Numerical Considerations}
\label{sec:numerics}

\begin{figure}
\sidecaption
\includegraphics[scale=0.29]{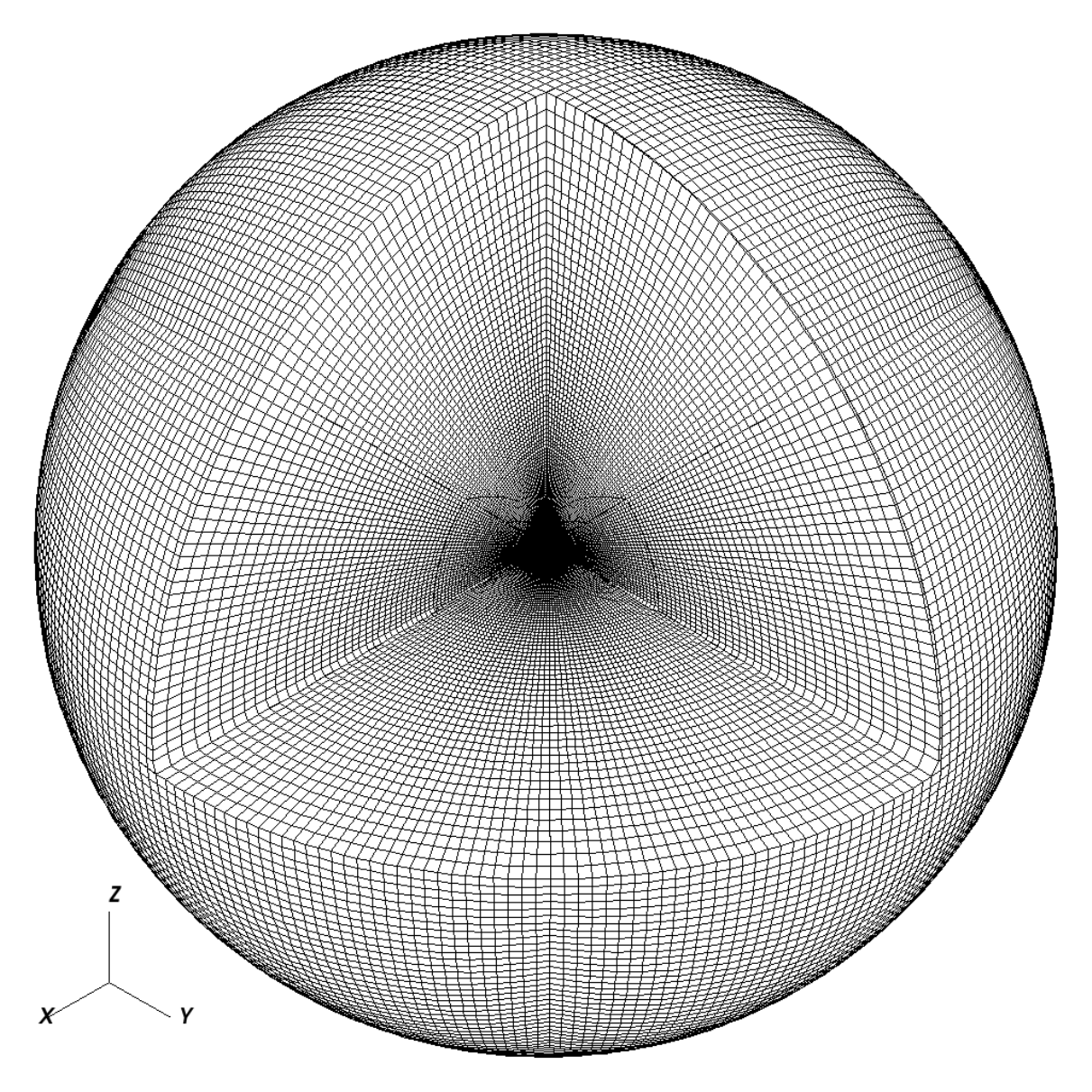}
\includegraphics[scale=0.29]{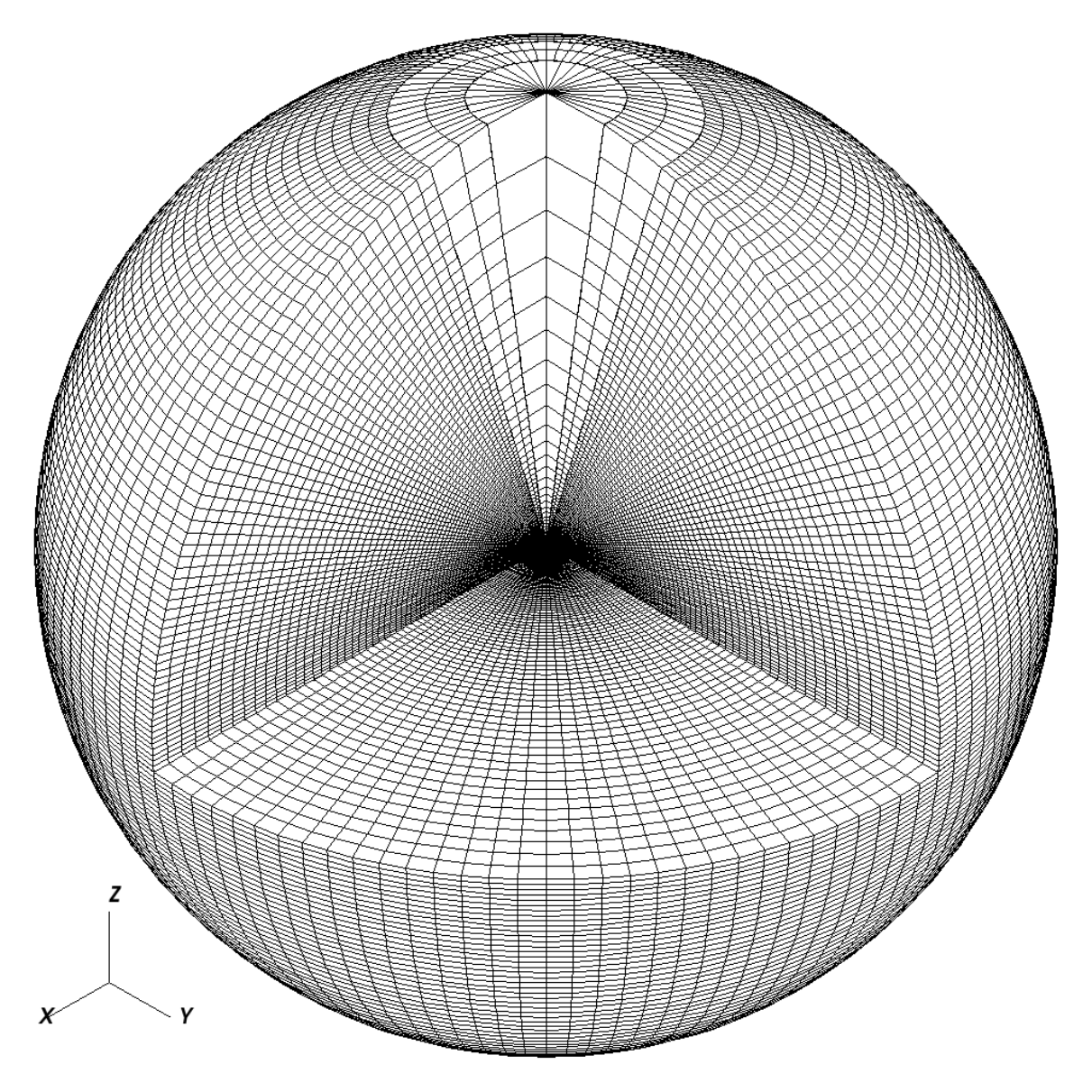}
%
%
\caption{Left: plot of the cubed-sphere grid geometry. Right: plot of a comparable spherical–polar grid with multiple resolution layers. Image reproduced by permission from \cite{Fragile09}, copyright by AAS.}
\label{fig:cubed_sphere}       
\end{figure}

Before proceeding with our review of the simulations themselves, we feel it may be worthwhile to mention a few of the special challenges associated with tilted disks. Tilted disk simulations are a level up in complexity in many ways when compared to their untilted counterparts. First off, tilted disks break any symmetries that may be exploited in simulations of untilted disks. This has two significant implications: 1) the simulations must always be done in full 3D, as there is no way to apply reflection or axial symmetry to reduce the computational cost; and 2) if one chooses to tilt the black hole with respect to the grid coordinates, then all the components of the resulting spacetime metric for gravity will be non-zero, which can potentially add computational cost relative to sparser, flat-space or symmetric metrics. A second challenge is choosing an appropriate grid and coordinate combination. Spherical-polar coordinates are commonly used in untilted black hole accretion simulations, which makes sense as such coordinates can be aligned with the predominant (azimuthal) direction of the flow to reduce numerical errors associated with advection oblique to cell faces. However, again, such symmetries are not as prevalent in tilted disks. Furthermore, the magnetohydrodynamic (MHD) equations are especially challenging to solve around the polar singularity. To minimize numerical artifacts when parts of the accretion disk pass through this artifact of the grid, transmissive polar boundary conditions can be used to minimize disturbances to the fluid \cite{McKinney13, Liska18, Liska20}. These allow the gas to pass through the polar singularity more naturally than the more commonly used reflecting boundary conditions, but they still induce excess dissipation. Thus, in cases of extreme tilt, especially when the disk passes through the pole, it may make sense to consider alternative coordinate and grid choices, such as the cubed sphere \cite{Fragile09} or even a Cartesian grid (see Figure \ref{fig:cubed_sphere} for a comparison of the cubed-sphere and spherical-polar grids).

\section{Simulations of Tilted Thick Disks}

\subsection{Discovery of Global LT Precession}
\label{sec:precession}

We begin our review of tilted disk simulations by considering those in the wave-like (thick-disk) limit, as this is numerically the easiest case and was, hence, the first one considered in general relativistic simulations. The early simulations all started with finite, rather small (tens of $r_G$), reservoirs of gas, usually in the form of a torus orbiting the black hole. These were largely numerical experiments, as there were very few published predictions for what should come out of such simulations. 

\begin{figure}
\sidecaption
\includegraphics[width=1.0\linewidth,trim=0mm 0mm 0mm 0,clip]{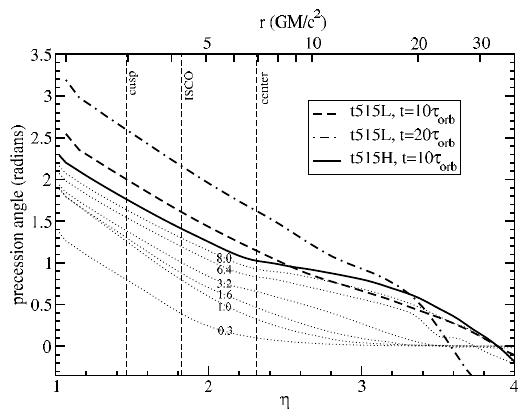}
%
%
\caption{Cumulative
precession angle $\gamma$ in radians as functions
of radius for times spanning from 0.3 to 20 orbital timescales for a tilted disk simulation with $\beta = 15^\circ$ and $a_* = 0.5$. Initially, $\gamma=0$ throughout the disk.  The bend in the profile around $6 r_G$ reflects the initial period of relatively rapid differential precession. However, once that profile is established, it remains largely unchanged as the disk transitions to more global precession. Image reproduced by permission from \cite{Fragile05}, copyright by AAS.}
\label{fig:precession}       
\end{figure}

One of the biggest surprises was the discovery of global precession of the torus following a brief period of differential twisting \cite{Fragile05}. A nice illustration of this behavior is provided in Figure \ref{fig:precession}, where the disk twist profile shows both differential precession (mostly at small radii) and steady growth of the total precession angle with time at all radii. Similar results have now been confirmed in multiple general relativistic magnetohydrodynamic (GRMHD) studies \cite{Fragile07, Fragile08, Liska18, White19}.

In retrospect, it seems obvious that global precession must occur whenever the sound-crossing time in the accretion flow is shorter than the precession timescale, which it almost always is for compact tori and disks. In fact, earlier Newtonian simulations had already pointed to this behavior \cite{Larwood96, Nelson00}. We can even write down an expression for the rate of disk precession. Generically, the precession period for a solid-body rotator with angular momentum $\mathbf{J}$ subject to a torque $\tau$ is simply
$T_{\rm prec} = 2\pi (\sin \beta) (J/\tau)$ \cite{Liu02}, where $\beta$ is the tilt angle between $\mathbf{J}$ and the black hole spin axis. The important point about this is that the LT torque gets integrated over the entire disk, so it has a cumulative effect. To illustrate, if we assume a power-law radial dependence to the surface density of the form $\Sigma \propto r^{-\zeta}$ and ignore higher order general relativistic corrections, we can write that
\begin{equation}
T_{\rm prec} = \frac{\pi (1+2\zeta)}{(5-2\zeta)}
\frac{r_o^{5/2-\zeta} r_i^{1/2+\zeta} \left[1-(r_i/r_o)^{5/2-\zeta}
\right]} { aM \left[1-(r_i/r_o)^{1/2+\zeta}\right]} 
\label{eqn:precession}
\end{equation}
(in units of $G=c=1$), where $r_i$ and $r_o$ are the inner and outer radii of the disk, respectively. This quantitative prediction was first confirmed in the simulations of \cite{Fragile07}. Interestingly, it was pointed out that for small disks, or isolated disk components, with $r_o \approx 10 r_G$, this precession period falls in the proper range to match what is known as the type-C class of quasi-periodic oscillations (QPOs) seen in the X-ray light curves of BHBs \cite{Ingram09}. We elaborate further on the potential association of QPOs with tilted disks in Section \ref{sec:qpos}.

\subsection{Unique Structural Features of Tilted Thick Disks}
\label{sec:standing_shocks}

\begin{figure}
\sidecaption
\includegraphics[scale=.5]{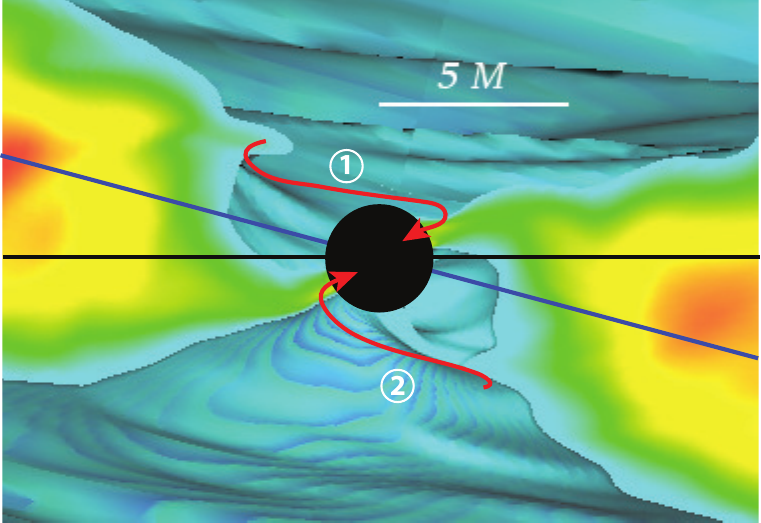}
\includegraphics[scale=.25]{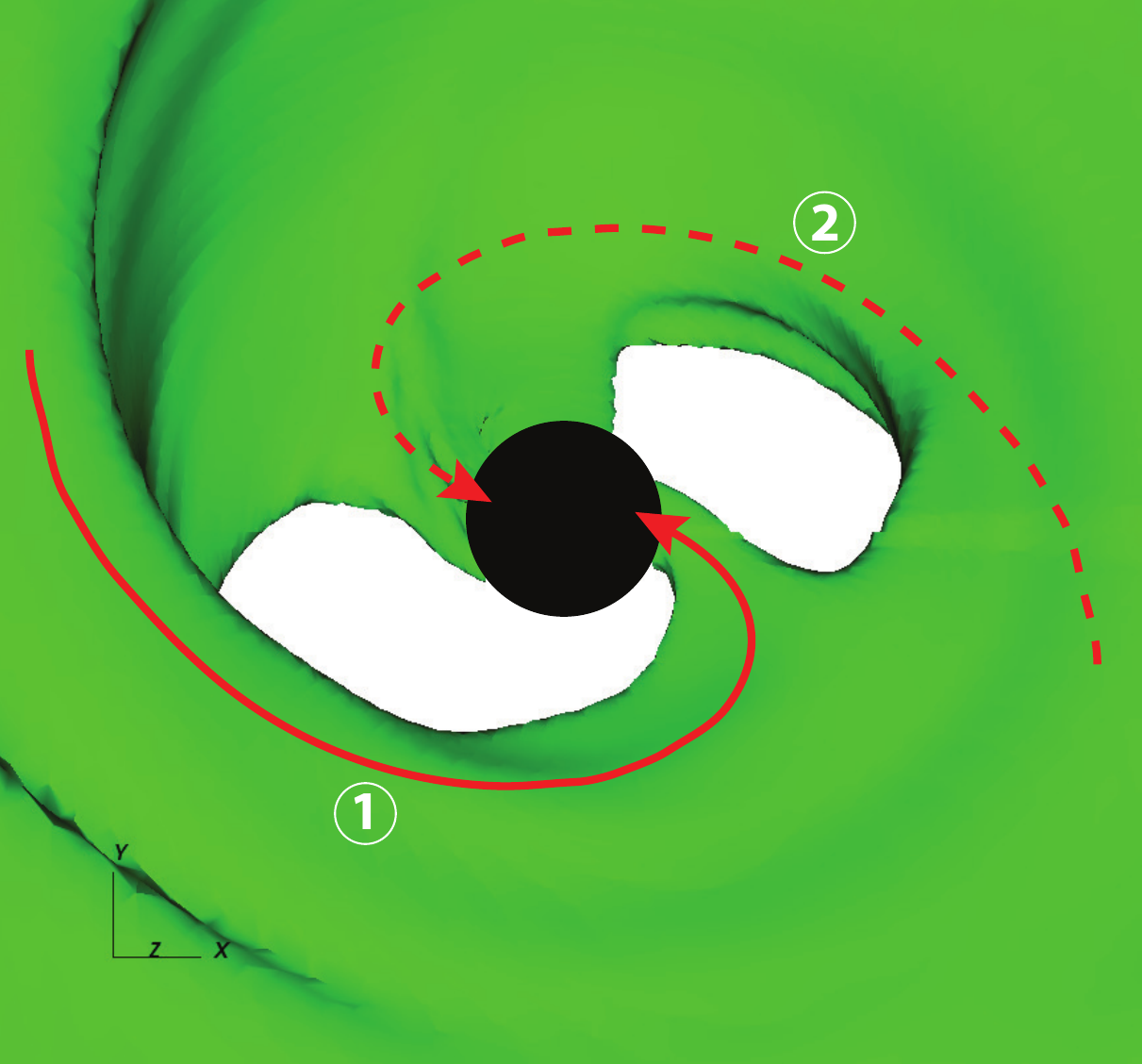}
%
%
\caption{Left: Cut-away view of the inner $10 r_G$ of a tilted accretion flow revealing two opposing, high-latitude features connecting the disk to the horizon (red arrows). The figure is oriented with the black hole spin ($a_*=0.9$) axis vertical. The black hole symmetry plane (black line) and initial disk midplane (blue line) are marked for reference. Right: Isodensity contour viewed down the spin axis of the black hole, highlighting the same two features as the left panel. Note that feature (1) remains entirely above both the black hole symmetry plane and disk midplane, while feature (2) remains below them. Images reproduced by permission from \cite{Fragile07}, copyright by AAS.}
\label{fig:streams}       
\end{figure}

Early tilted thick-disk simulations also revealed interesting non-axisymmetric structures, particularly in the innermost regions of the disk, that had not been seen in comparable simulations of untilted disks. One example is the pair of over-dense bridges of matter appearing to connect the main body of the disk to the black hole, as seen in Figure \ref{fig:streams}. These features were originally referred to as ``plunging streams'' \cite{Fragile07}, although it was later shown that they are better described as post-shock, over-density features, since the actual fluid trajectories are roughly normal to these structures, as opposed to along them \cite{Generozov14}. Regardless, it is remarkable that these over-dense bridges (and the intervening low-density pockets) are {\em persistent} features, lasting the entire duration of the simulations and even precessing along with the rest of the inner disk \cite{Fragile08}. 

Since the persistent, over-dense bridges are actually post-shock enhancements, this means that the shocks that create them are {\em standing} shocks, having fixed orientations with respect to the line-of-nodes between the disk and black hole midplanes and also persisting for the duration of current simulations. Standing shocks could profoundly affect the behavior of tilted disks by enhancing the angular momentum transport, increasing the mass accretion rate, increasing the energy dissipation, changing the effective innermost radius of the disk \cite{Fragile09b}, and accelerating particles \cite{Sironi24}. 

\begin{figure}
\sidecaption
\includegraphics[width=1.0\linewidth,trim=0mm 0mm 0mm 0,clip]{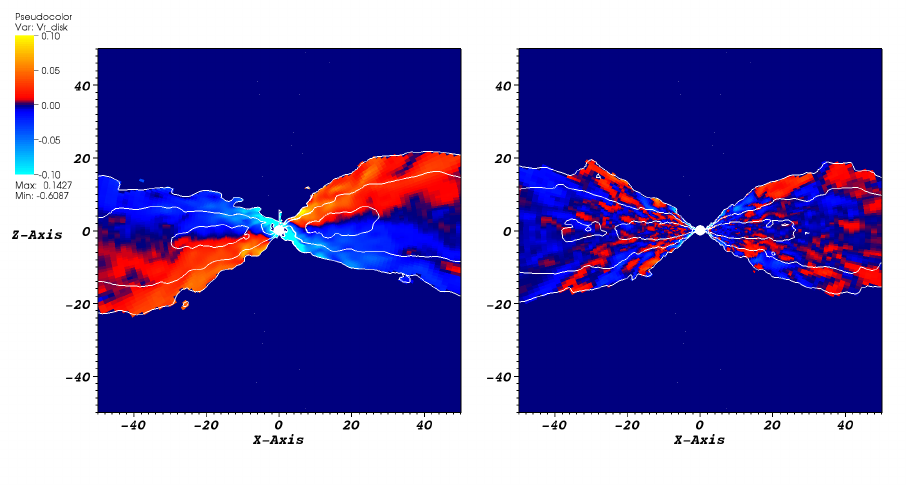}
%
%
\caption{Meridional ($\phi=0$) plots for tilted ({\em left}) and untilted ({\em right}) simulations showing a pseudocolor representation of $V^r$ for outflowing ($V^r > 0$) and inflowing ($V^r < 0$) gas as {\em hot} and {\em cold} colors, respectively. The velocity scale is normalized to the speed of light $c$. The plots are overlaid with isocontours of density at $\rho = 0.4$, 0.04, and $0.004\rho_\mathrm{max,0}$. The figure in the {\em left} panel is oriented such that the black hole is tilted $15^\circ$ to the left. Note the pattern of radial epicyclic motion in the image on the {\em left} that is clearly absent from the image on the {\em right}. Image reproduced by permission from \cite{Fragile08}, copyright by AAS.}
\label{fig:Vr}       
\end{figure}

Further study revealed that the non-axisymmetric features owe their existence to the fact that orbital trajectories are more eccentric in tilted disks than untilted ones, and the eccentricity rises strongly with decreasing radius, especially inside $15 r_G$ (see Fig. 16 of \cite{Dexter11}, for example). This rise in eccentricity with decreasing radius inevitably leads to a crowding of orbital trajectories near their respective apocenters \cite{Ivanov97}. Such crowding of the orbital trajectories provides the necessary conditions for standing shocks to arise. The reason there are {\em two} standing shocks, located on opposite sides of the black hole with one above the midplane and one below, is because the eccentricity of the orbital trajectories is 180$^\circ$ out of phase across the disk midplane. In other words, the apocenters of the orbits above the midplane come together on the opposite side of the black hole from where they come together below the midplane. Another way to picture this is in terms of the radial epicyclic motion of the gas in the disk. Figure \ref{fig:Vr} shows material moving radially outward on one side of the black hole and radially inward on the other side (indicative of eccentric trajectories), with the pattern being reversed across the disk midplane (left panel). Similar behavior is not seen in an equivalent untilted simulation (right panel).

\subsection{SANE vs. MAD}
\label{sec:MAD}

Before discussing results for thin disks and jets, we first need to distinguish between two general types of black hole accretion simulations: standard and normal evolution (SANE) \cite{Narayan12} and magnetically arrested disks (MAD) \cite{Narayan03}. Most of the early work, for both untilted and tilted disks, focused on the weakly magnetized (SANE) case, with MADs becoming more popular in recent years. This has also begun to translate into the realm of tilted disks as well. The thick disks that we have discussed so far that precess globally \cite{Fragile07, Liska18, White19} were all SANE disks. 

However, for MADs, the case can be very different. MADs are formed when the accretion disk is saturated by vertical magnetic flux, and the corresponding magnetic pressure comes within a factor of a few of equipartition with the gas pressure. Whenever this happens, the disk attempts to get rid of the excess magnetic flux by rapidly ejecting flux bundles toward larger radii, allowing the magnetic pressure to drop. The radius out to which the disk is saturated, which we will call $r_\mathrm{MAD}$, is set by, among other things, the amount of vertical magnetic flux in the system. In some MADs, the disk will only be saturated very close to the black hole, while in other cases, excess magnetic flux might be present out to hundreds, if not thousands, of $r_G$. The plasma within $r_\mathrm{MAD}$ typically has a much larger radial infall speed, and, because of mass continuity, a correspondingly lower density than the plasma beyond this radius. While many numerical studies include MAD and SANE models, it is important to note that not all MAD or SANE models are the same. Each can experience a wide range of magnetic flux values. Even the boundary between MAD and SANE is merely determined by the saturation value of the magnetic flux at the event horizon. There can be remarkable differences between SANE states that are, for example, just below the saturation level and SANE states that contain no vertical magnetic flux at all. Likewise, MADs will depend on the total amount of excess magnetic flux in the disk, in other words on the value of $r_\mathrm{MAD}$.

Going back to the question of precession, simulation results obtained so far suggest that the disk {\em does not} precess when the accretion is MAD \cite{McKinney13, Ressler23, Chatterjee23}. This seems to be because, in the MAD state, the black hole ergosphere and inner accretion disk are saturated with magnetic flux, which, as it is twisted by the rotating black hole, is forced to align with the black hole spin axis. Because the dynamics of the inner disk are dominated by the magnetic field in the MAD state, the inner disk aligns in response to the alignment of the field. In other words, the tilt of the inner disk is erased by the effects of the MAD field. Even for initial tilts up to $60^\circ$, MADs are seen to be forced to align with the symmetry plane of the black hole out to $r \gtrsim 10 r_G$ \cite{Chatterjee23}. This forced alignment obviously prevents LT precession of the disk. However, we caution that all numerical work considering tilted MADs so far has used rather large values of $r_\mathrm{MAD}$, meaning more work needs to be done on tilted MADs.

\section{Simulations of Tilted Thin Disks}

\subsection{Status of the Bardeen-Petterson Effect}
\label{sec:BPeffect}

The BP effect is often quoted as one of the reasons why tilt might actually not be an important parameter for accretion disks. The logic is that if the BP effect aligns the disk with the black hole out to a relatively large radius (100-1000 $r_G$), then the disk can, for all intents and purposes, be treated as an untilted disk. The trouble with this thinking is that the validity of the BP effect, at least based on numerical studies, is somewhat ambiguous to say the least. On the one hand, smoothed-particle hydrodynamic (SPH) simulations \cite{Nelson00, Lodato07, Lodato10} have largely confirmed the analytic predictions of \cite{Papaloizou83}, but were subject to the same assumptions, namely that the physics is adequately described by Newtonian gravity plus a torque term, that the ``viscosity'' is isotropic, and that the Shakura \& Sunyaev \cite{Shakura73} $\alpha$-parameter is constant. However, each of those assumptions is known to be incorrect \cite{Sorathia13, Teixeira14}.

\begin{figure}
\sidecaption
\includegraphics[width=1.0\linewidth,trim=0mm 0mm 0mm 0,clip]{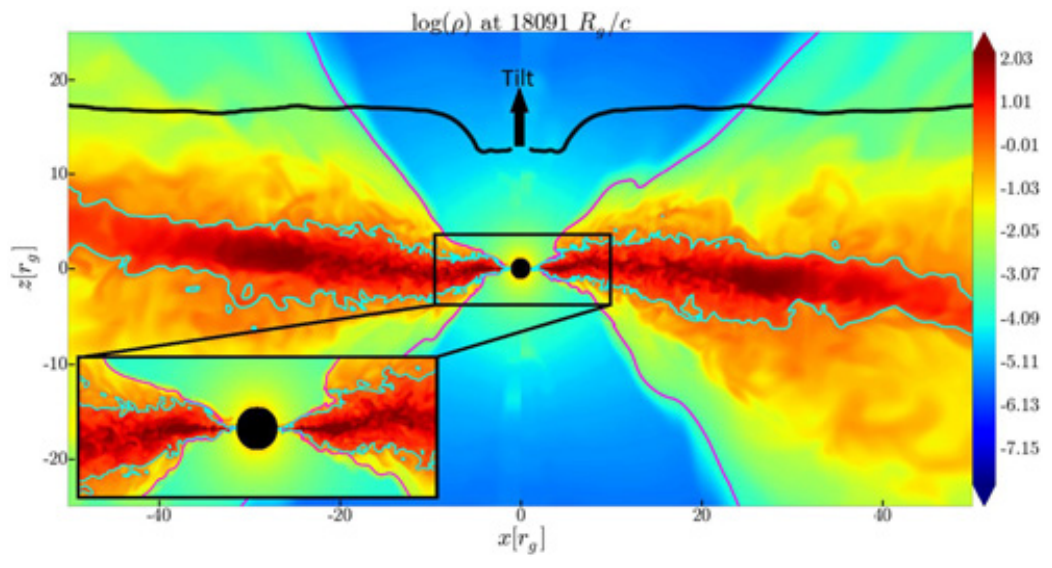}
%
%
\caption{Pseudocolor plot of density for a thin ($\delta = 0.03$), tilted disk, showing that only the innermost part of the disk ($r \lesssim 5 r_G$) has aligned with the black hole symmetry plane (horizontal in this image). Image reproduced by permission from \cite{Liska19}, copyright by RAS.}
\label{fig:BPLiska}       
\end{figure}

On the other hand, it has taken a while for full GRMHD simulations to weigh in on the BP effect. With grid-based codes, tilted thin-disks are very challenging because of the difficulty in resolving all three dimensions. As an example, if you want 20 zones to cover the disk vertical scale height $\delta$, then to have comparable resolution in the radial and azimuthal directions would require something like $80/\delta$ and $40\pi/\delta$ zones, respectively. For $\delta \approx 0.1$, that is approaching a $1000^3$ computational domain, which is still challenging for all but the largest supercomputers. Nevertheless, \cite{Teixeira14} probed the regime where $\alpha \approx \delta \lesssim 0.1$, i.e., when the BP effect should begin to apply. They found that their prograde tilted disks showed {\em no sign} of BP alignment even in the innermost regions. Only their retrograde case showed even modest evolution toward anti-alignment inside $10 r_G$. \cite{Liska19} pushed to even thinner disks ($\delta \approx 0.015-0.03$) and saw modest BP alignment in prograde disks out to $\sim 5-10 r_G$ (an illustration is provided in Figure \ref{fig:BPLiska}). The implication of these results is that the parameter space associated with the BP effect may be rather small, only including very thin disks, and even so, the alignment appears to be much more modest than what was originally predicted in \cite{Bardeen75}.

\subsection{Disk Tearing}
\label{sec:tearing}

In Section \ref{sec:precession}, we discussed precession in the context of relatively small, isolated, thick accretion disks (few 10s of $r_G$). The precession period thus found was in a range consistent with the frequency of QPOs in astrophysical BHBs. However, accretion disks in real, physical systems are orders of magnitude larger than those used in early simulations. Since the Lense-Thirring precession rate decreases proportional to $r^{-3}$, we would thus expect such large accretion disks to precess at a much lower frequency due to their immense size (and much larger total angular momentum).

\begin{figure}[t!]
\begin{center}
\includegraphics[width=1.0\linewidth,trim=0mm 0mm 0mm 0,clip]{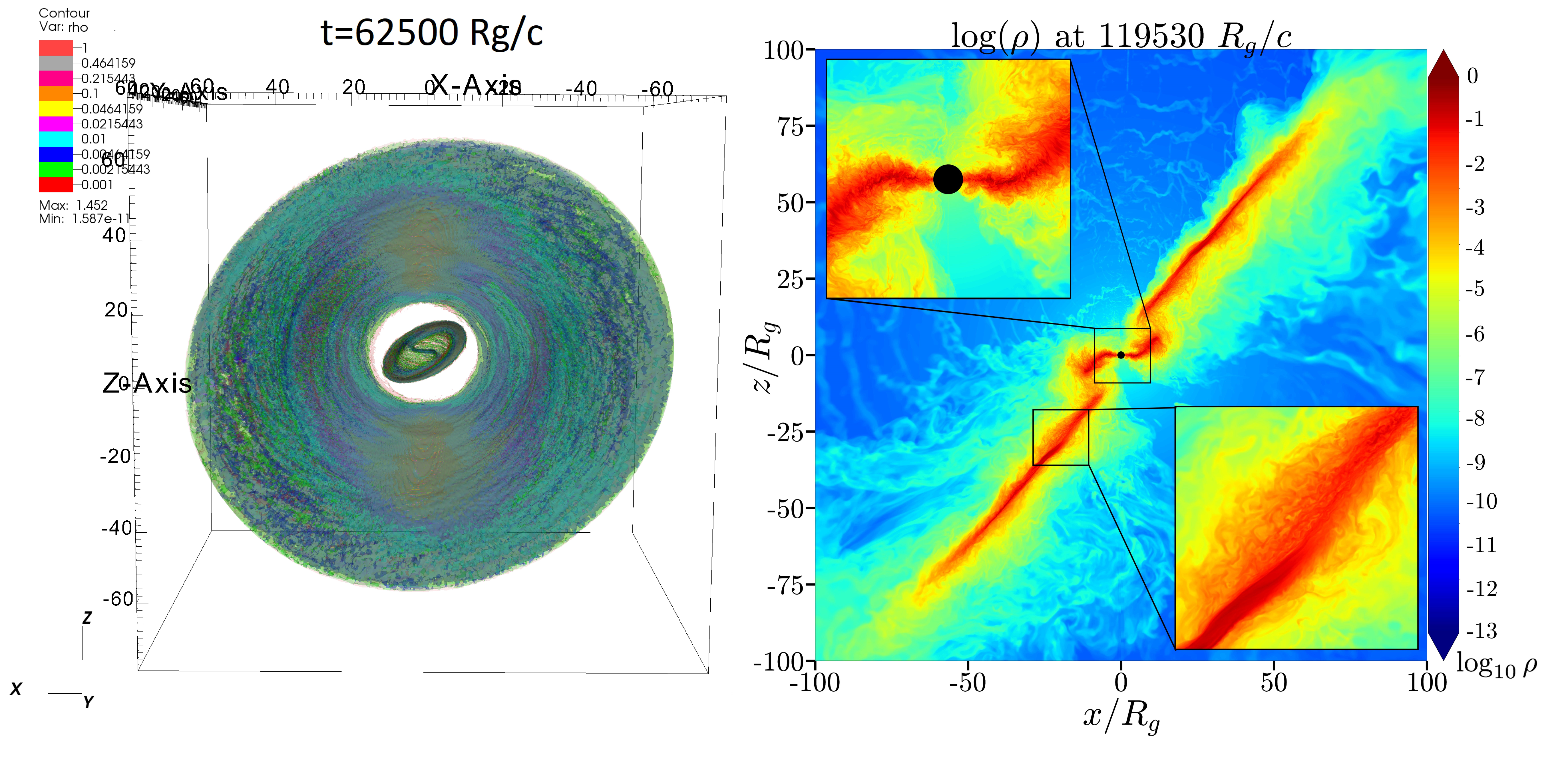}
\end{center}
\vspace{-0.8cm}
\caption{An extreme resolution GRMHD simulation \cite{Liska20, Musoke23} of a thin accretion disk (aspect ratio $\delta=0.02$), which tears apart into an outer, non-precessing disk and inner, precessing one. The initial disk tilt is $\beta = 65^{\circ}$ relative to the (horizontal) equatorial plane of the black hole. The right panel shows a vertical slice through the density with the black hole spin pointing vertically. Image reproduced by permission from \cite{Liska20}, copyright by AAS.}
\label{fig:tearing}
\vspace{-0.6cm}
\end{figure}

One way to circumvent this problem, and recover a reasonable precession period, would be to tear off a smaller, precessing sub-disk from the larger, non-precessing parent disk. This process, called disk tearing, was first demonstrated in SPH simulations \cite{Nixon12b, Nealon15} in which disks governed by an isotropic $\alpha$-viscosity tore apart into multiple independently precessing rings. Later, GRMHD simulations confirmed that disks can tear \cite{Liska19b, Liska20}, though they have so far found the disk to only tear off into a single, radially extended, precessing sub-disk rather than a multitude of individual rings (see Figure \ref{fig:tearing}). The reason for this discrepancy is still an area of investigation, but magnetic torques \cite{Liska19b}, which are not included in SPH simulations, may be a key difference. 

The exact location of the tearing radius is difficult to predict. Naively, one might expect a disk to tear if the viscous torques within the disk are not strong enough to counter the differential Lense-Thirring precession of the surrounding spacetime. In other words, if we write the Lense-Thirring torque as $\tau_{LT}$ and the viscous torque as $\tau_{\nu}$, then tearing may be expected whenever $\tau_{LT} \gtrsim \tau_{\nu}$. The radius where this criterion is met, or the so-called tearing radius $r_\mathrm{tear}$, can be written \cite{Nixon12b}
\begin{equation}
r_\mathrm{tear} \lesssim \frac{4}{3}| \sin \beta|\frac{a}{\alpha\delta} ~.
\label{eqn:tearing_radius}
\end{equation}
However, it turns out that this expression does not agree well with the results of GRMHD simulations \cite{Liska19b}. This could be because the viscosity in these simulations is not isotropic, as it is assumed to be in analytic work and SPH simulations, or because the viscosity coefficient is not constant throughout the disk. The behavior of $\alpha$ in regions of large warp amplitude is also poorly understood. What is known is that when the LT torque is increased, the disk needs to compensate by increasing the warp amplitude to remain a rigid body. Typically, such an increase in warp amplitude would also lead to an increase in the viscous torques responsible for transporting misaligned angular momentum. However, some semi-analytical work suggests that, at a critical warp amplitude, the viscous torque becomes weaker instead of stronger, allowing the disk to tear apart \cite{Dogan18}.

\subsection{Nozzle Shocks}

\begin{figure}
    \includegraphics[width=1.0\linewidth,trim=0mm 0mm 0mm 0,clip]{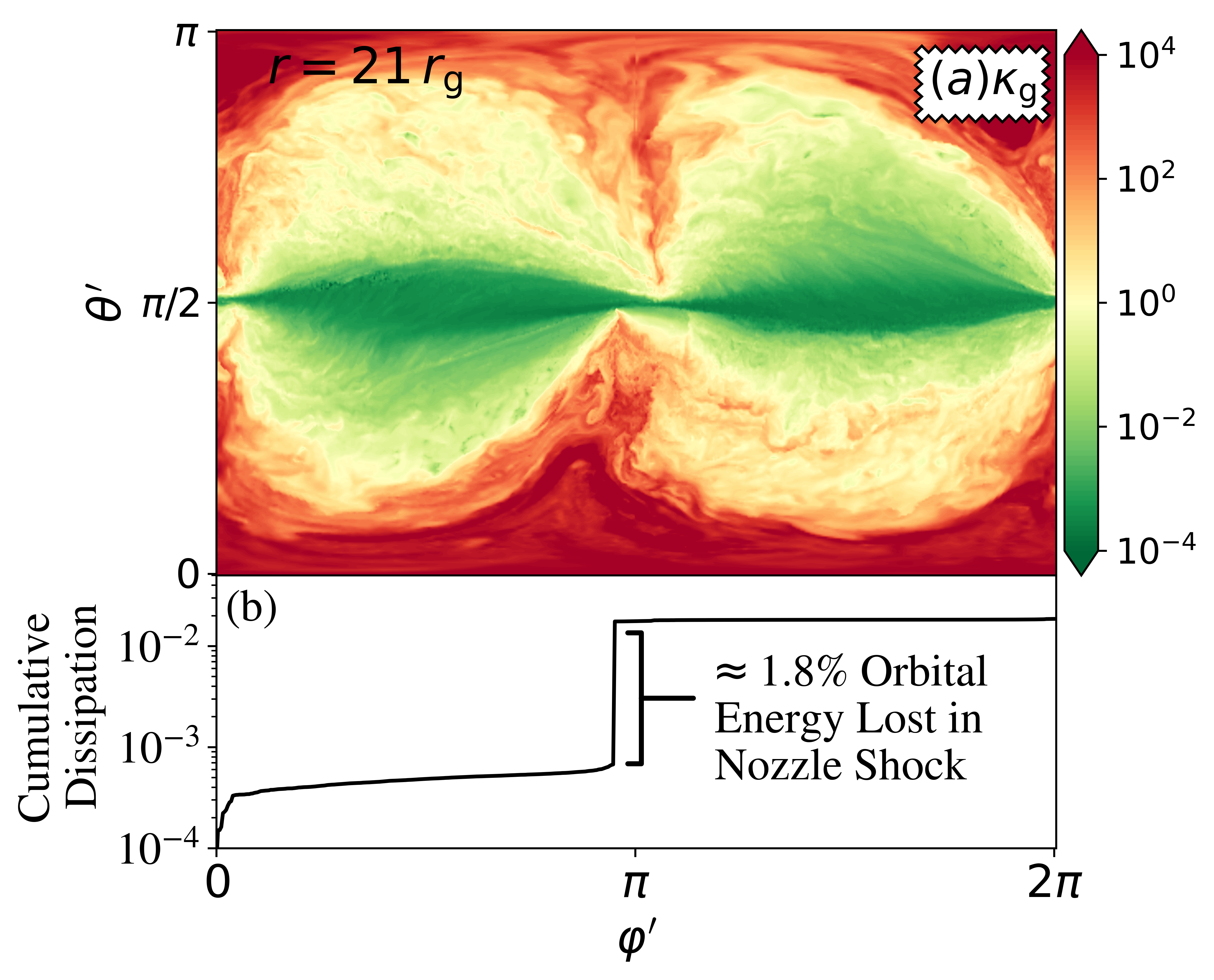}
    \caption{Vertical compression in simulations of warped, thin disks leads to nozzle shocks at two azimuths. Panel (a): Fluid-frame entropy ($\kappa_{\rm g}$), at radius $r=21 r_G$ in the $\theta^\prime-\varphi^\prime$ plane showing spikes at $\varphi^\prime\approx0$ (or $2\pi$) and $\pi$, where the disk is most compressed. Panel (b): The cumulative fraction of dissipated energy along the annulus, normalized to the orbital energy. Across the $\varphi^\prime=\pi$ nozzle shock, $\approx1.8\%$ of the orbital energy is dissipated. Image reproduced by permission from \cite{Kaaz22}, copyright by AAS.}
    \label{fig:flowNozzleEntropy}
\end{figure}

As described in Section \ref{sec:standing_shocks}, the increasing eccentricity of particle orbits towards smaller radii in warped accretion disks can lead to a crowding that produces standing shocks. In thick disks with modest tilts, these standing shocks have only a modest effect on the accretion rate \cite{Fragile07}. However, in thin disks with large tilts, the disk undergoes vertical compression twice per orbit (see Figure \ref{fig:flowNozzleEntropy}) that is strong enough to produce nozzle shocks, which increase the radial infall speed of the gas by 2-3 orders of magnitude \cite{Kaaz22}. Whenever a particle passes through one of the nozzle shocks, it can lose several percent of its orbital kinetic energy, leading to an effective $\alpha$ of up to $10^2-10^3$, possibly providing an explanation for the rapid variability seen in certain changing-look AGN. 

\label{sec:accretion:nozzleshocks}

\begin{figure}
\includegraphics[width=1.0\linewidth,trim=0mm 0mm 0mm 0,clip]{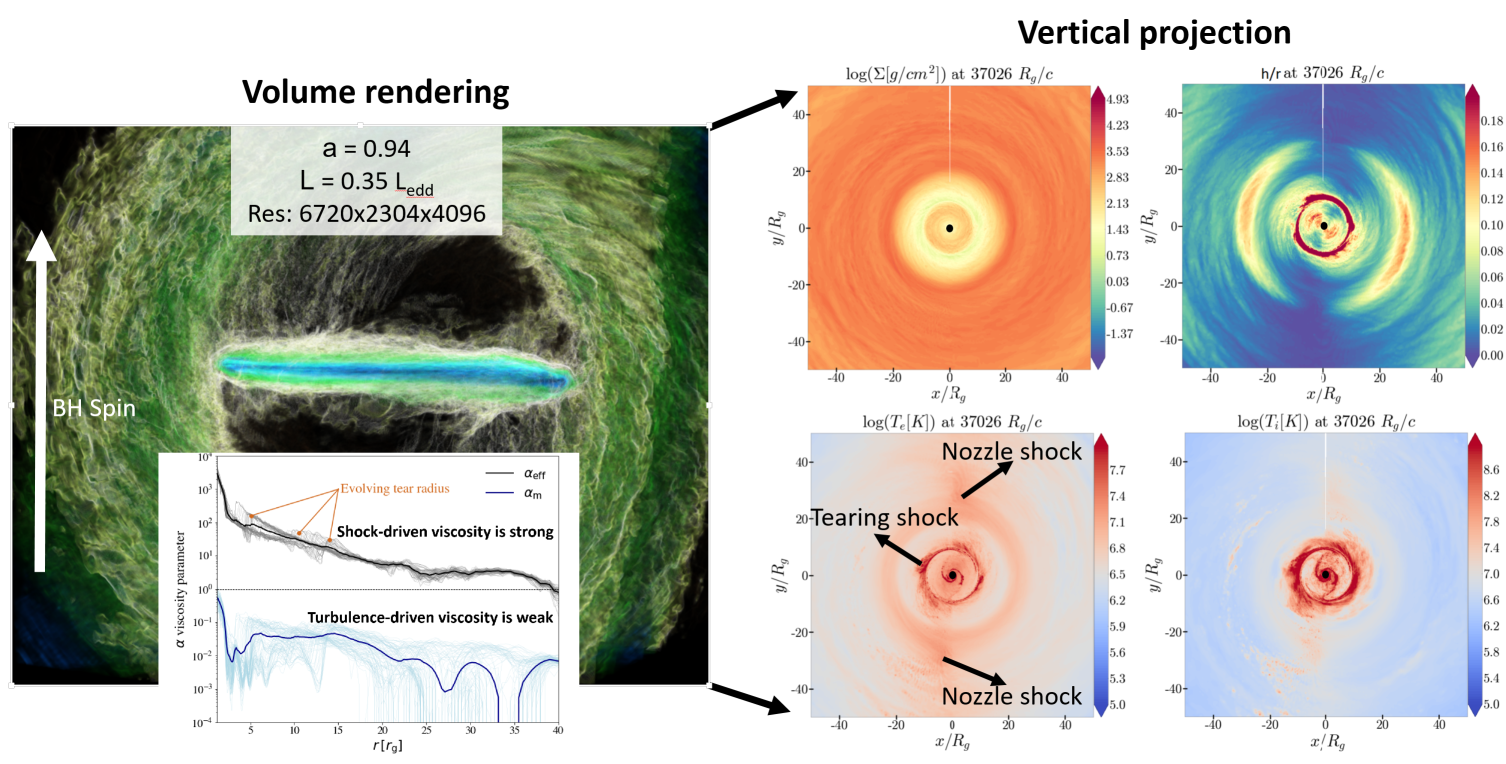}
\caption{Volume rendering of a radiative GRMHD simulation of a misaligned accretion disk in the ``high-soft'' state (left panel) alongside the vertically integrated density $\Sigma$, scale height $h/r$, electron temperature $T_e$, and ion temperature $T_i$ (right panel). In this accretion disk the warping of space-time causes the disk to tear apart into two sub-disks: a precessing, inner disk and a non-precessing, outer one. It demonstrates that shock dissipation occurs at two locations, where it leads to a substantial increase in the electron and ion temperatures: 1) Perpendicular to the line of nodes along the z-axis; and 2) at the tearing radius. The inset in the left panel demonstrates that the shock-induced effective viscosity ($\alpha_{\rm eff}$) is orders of magnitude larger than the turbulent viscosity seeded by MRI turbulence ($\alpha_{\rm m}$). Image produced by author from data published in \cite{Kaaz22, Liska23b}, copyright by AAS.}
\label{fig:shocks}
\end{figure}

More recent two-temperature, radiative GRMHD simulations \cite{Liska23b}, which treated the ions and electrons as separate fluids and solved for the radiation, found that nozzle shocks also change the structure and spectral signatures of disks (see Figure \ref{fig:shocks}). Namely, the infalling gas follows a non-axisymmetric temperature profile with hotspots centered around the nozzle shocks and the tearing radius. While the heating of the gas in the nozzle shocks is mostly driven by adiabatic compression, as the scale-height decreases by an order of magnitude within the nozzle, gas crossing the tearing radius heats up dramatically as it undergoes a rapid orbital plane change crossing from the outer, parent disk to the inner, sub-disk. 

\section{Simulations of Tilted Truncated Disks}
\label{sec:truncated}

In many systems we are not dealing simply with either a thin disk or a thick one, but with a hot, thick inner flow coupled to a cold, thin outer disk at a ``truncation radius'' \cite{Esin97, Ferreira06, Done07}. Such a truncated disk picture is typically used to explain the hard state of BHBs whenever $L \gtrsim 10^{-3} L_{\rm Edd}$. Truncated disks naturally produce a spectrum that is dominated by Compton-scattered X-ray emission on top of a weaker, thermal blackbody component. As of the time of this writing, the physics behind disk truncation is still poorly understood. The best known models invoke thermal conduction from the corona to the disk leading to evaporation \cite{Meyer94, Liu02, Qian07, Cho22}. However, recent shearing-box MHD simulations \cite{Bambic24} suggest that the plasma may cool too quickly to be able to efficiently conduct heat, thus raising doubt about the feasibility of this picture.

\begin{figure}
\includegraphics[width=1.0\linewidth,trim=0mm 0mm 0mm 0,clip]{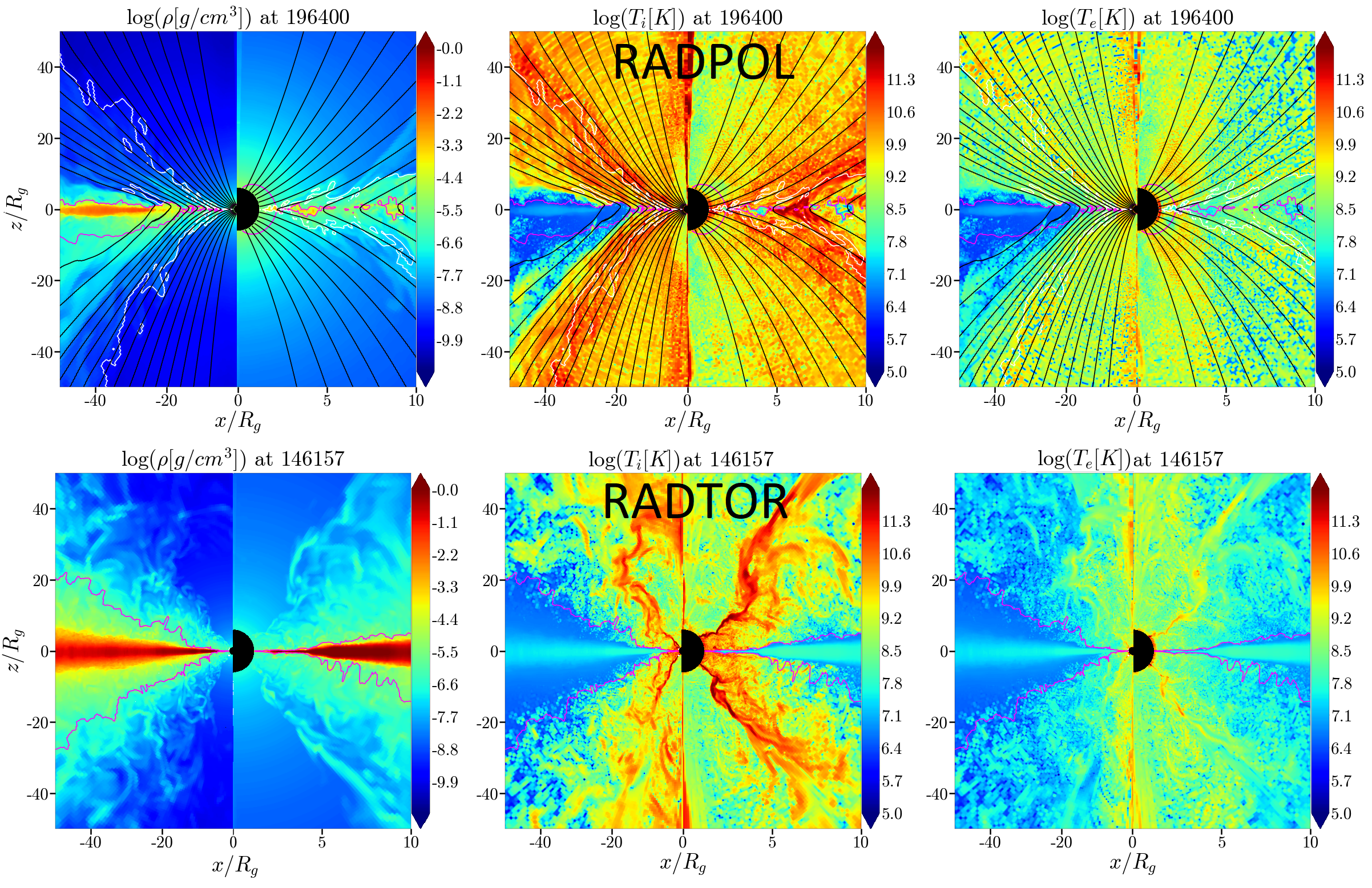}
\caption{The presence of large-scale poloidal magnetic flux leads to the development of a two-phase medium: a low-density, thick, hot corona-like accretion flow with clumps of cold gas floating through it (top row). In contrast, in the absence of the poloidal magnetic flux, the cold, thin flow extends down to the black hole (bottom row). The three columns, from left to right, show vertical slices through the density, ion temperature, and electron temperature; the right hemispheres show zoom-ins on the left hemispheres. Magnetic field lines are shown in black, jet boundary (${p_{mag}}={\rho}c^2$) in white, and the last scattering surface ($\tau_{es}=1$, integrated in the vertical direction) in pink. Image reproduced by permission from \cite{Liska22}, copyright by AAS.}
\label{fig:truncated_corona}
\end{figure}

As an alternative, recent radiative, two-temperature GRMHD simulations \cite{Liska22} have suggested that truncation might instead be set by $r_\mathrm{MAD}$. In these simulations, the saturation of magnetic flux in the inner disk can cause it to become magnetic-pressure-dominated, allowing reconnection layers to effectively heat up the gas to extreme temperatures, consistent with a hot corona. Interestingly, within this hot corona were found cold clumps of much denser gas (see Figure \ref{fig:truncated_corona}). These cold clumps might explain the presence of broadened iron lines seen in the hard state \cite{Reis10}, which are difficult to explain in the standard truncated disk picture.

Sidestepping the question of how truncated disks might form, recent GRMHD simulations \cite{Bollimpalli23} investigated the question of precession in tilted, truncated disks by simply prescribed a radially dependent scale-height for the disk using a cooling function. They confirmed that the corona region inside the truncation radius can precess as a rigid body independent of the outer disk, though possibly at a lower frequency than an isolated corona. This is in line with semi-analytical work \cite{Marcel2020} suggesting there might exist some correlation between the frequency of precession-induced type-C QPOs and the disk truncation radius. 


\section{Jets from Tilted Disks}
\label{sec:jets}

Accretion disks throughout nature are very often associated with ``jets'' -- narrow streams of fast-moving plasma usually ejected perpendicular to the disk. Studying these jets has been a popular goal of black hole accretion simulations \cite{McKinney06, Tchekhovskoy11, Chatterjee19}. In this section, we look particularly at the case of jets from tilted disk simulations. Naively, one might expect that jets powered by the Blandford-Znajek process \cite{Blandford77}, which tap into the spin of the black hole, should align with the black hole spin axis, while jets powered by the Blandford-Payne process \cite{Blandford82}, which tap into the spin of the disk, should precess along with the disk. Current simulation results, though, present a more nuanced picture.

\subsection{SANE vs. MAD (revisited)}

\begin{figure}
\includegraphics[width=1.0\linewidth]{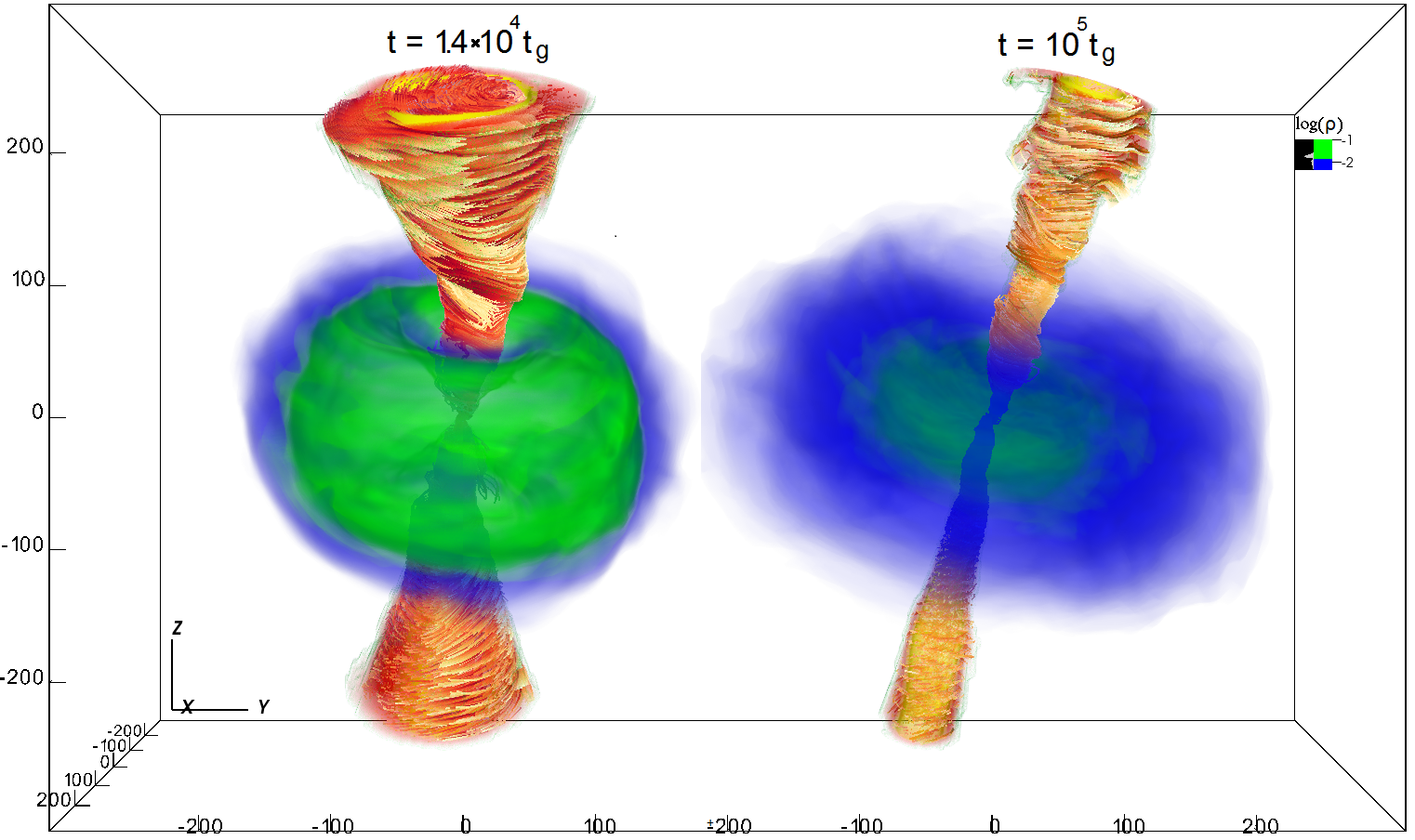} 
\caption{ Volume rendering of density [$\log(\rho)$ in blue and green, indicating the disk] and magnetic energy density [$p_B r / \rho c^2 > 0.5$ in red and yellow, indicating the jets] at early (left) and late times (right). The disk-jet system precesses as a whole around the black hole spin vector, which is vertical in the figure. 
Image reproduced by permission from \cite{Liska18}, copyright by RAS. }
\label{fig:precessing_jet}
\end{figure}

In the SANE case, where as we already mentioned in Section \ref{sec:precession}, the disks precess globally, the corresponding jets also precess {\em in phase} with the disk \cite{Liska18, Liska23, Ressler23}. In other words, the jet aligns with, {\em and follows}, the angular momentum axis of the disk, not the black hole spin axis. An example of this is shown in Figure \ref{fig:precessing_jet}. This applies for SANE disks with vertical magnetic flux up to the MAD level and is true even though these jets are powered by the Blandford-Znajek process. This apparent paradox, that the jet seems to be aligned with the accretion disk even though it is powered by the black hole, can be explained by the torque exerted by the disk-driven winds on the jets. These winds are responsible for collimating and orienting the jets. Because the winds are driven by the disk, they orient the jet according to the orientation of the accretion disk. In simulations at least, it appears that accretion disks that have sufficient magnetic flux to launch meaningful jets also launch powerful winds, which are naturally aligned with the accretion disks and can affect the direction of the jet. 

\begin{figure}
\includegraphics[width=1.0\linewidth,trim=0mm 0mm 0mm 0,clip]{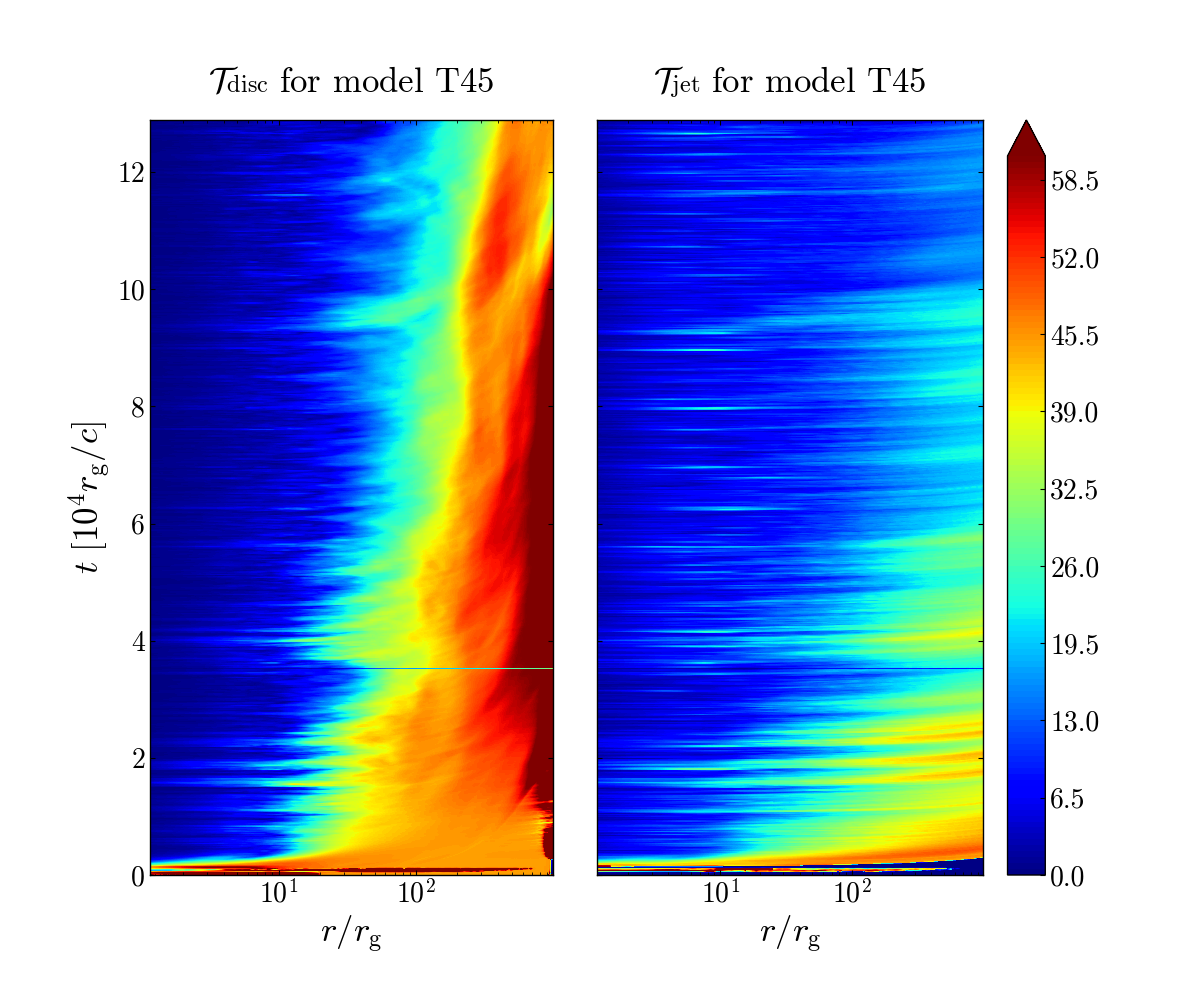}
\caption{Spacetime plots of the disk (left panel) and jet (right panel) tilt angles (in degrees) for model T45 of \cite{Chatterjee23}. The disk quickly aligns (tilt goes to $\approx$0) to beyond a radius of $10 r_G$, while the jet is also closely aligned with the black hole spin axis to beyond $10^2 r_G$. Image reproduced by permission from \cite{Fragile23}, copyright by RAS.}
\label{fig:jet}
\end{figure}

However, as we mentioned in Section \ref{sec:MAD}, published MAD simulations {\em do not} precess. This appears to be because the magnetic flux aligns with the black hole spin axis, which then forces the inner disk to align. Figure \ref{fig:jet} provides one example of this behavior (see also Table 1 of \cite{McKinney13}  and Fig. 14 of \cite{Ressler23} for further evidence). This forced alignment prevents LT precession of the disk, which, in turn, precludes any LT-driven precession of the jet. (Because the tilt is effectively $0^\circ$ for much of the jet, it is not even possible to define a reasonable precession angle.) Likewise, in cases where thin disks undergo BP alignment (Section \ref{sec:BPeffect}), there is also no precession of the jet, at least at small scales \cite{Liska19}.

Finally, we note that even in cases where the inner disk precesses (Sections \ref{sec:tearing} and \ref{sec:truncated}), some GRMHD simulations show that the jet at larger scales can be torqued into partial alignment with the outer, non-precessing disk \cite{Liska19b}. This suggests that it may be difficult to detect LT jet precession on scales beyond $r \gtrsim 10^2 r_G$. However, those particular simulations had relatively small accretion disks that were not continuously fed from larger scales. Future simulations that model larger accretion disks will be necessary to test whether or not jet precession can persist on scales order(s) of magnitude larger than the size of the inner precessing disk.


\section{Variability and Quasi-Periodic Oscillations}
\label{sec:qpos}

One of the earliest interests in tilted accretion disks stemmed from the possibility that the additional timescales associated with LT precession or the BP effect might correspond with one or more of the unexplained QPOs seen in the light curves of many accreting black hole systems \cite{Fragile01}. Although there is still much work to do to connect all the dots, the exciting news is that numerical simulations of tilted accretion disks have indeed demonstrated a richer, stronger, and more persistent set of timing signals than untilted disks, as we describe in this section.

\subsection{Low-Frequency QPOs}

The first ``timing signals'' that were observed in tilted disk simulations were those of LT precession itself, as already discussed in Section \ref{sec:precession}. It was recognized fairly early on that variations in the light curve from this type of precession would likely provide an excellent match to a particular type of low-frequency QPO called the type-C \cite{Ingram09, Ingram11}.\footnote{For a thorough review of QPO phenomenology in BHBs, see \cite{Ingram19}.} The strengths of this association are that: 1) it produces the correct range of QPO frequencies for reasonable black hole and disk parameters \cite{Ingram09, Motta18}; 2) it naturally yields a rise in the QPO frequency at the start of each outburst cycle when coupled with the truncated disk model \cite{Ingram11}; 3) it easily reproduces the correlation between system inclination and QPO strength \cite{Heil15, Motta15}; and 4) it also explains why the centroid energies of reflection features seen in phase-resolved spectroscopy vary with QPO phase \cite{Ingram16}. Together, these strengths make LT precession one of the most successful QPO models out there.

\subsection{High-Frequency QPOs}

Tilted disk simulations have also yielded some promising candidates for ``high-frequency'' QPOs ($\gtrsim 60$ Hz). For instance, \cite{Henisey09} found evidence for excess power around $118 (M/10 M_\odot)^{-1}$ Hz over an extended radial range in one early tilted-disk simulation. Later analysis revealed this to be a transient feature related to the orbit of an over-dense clump passing through the standing shocks \cite{Henisey12}; however, this effort was still important for opening the doors for later discoveries.

\begin{figure}
\sidecaption
\includegraphics[width=1.0\linewidth,trim=0mm 0mm 0mm 0,clip]{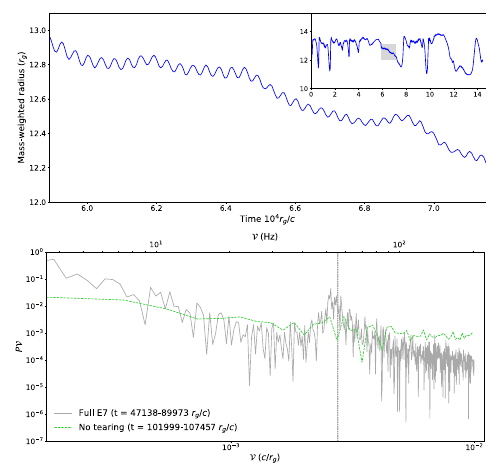}
%
%
\caption{Top: Radial oscillations of the inner sub-disk in a tearing-disk simulation. The inset panel shows the mass-weighted radius for the entire simulation duration, while the main panel shows a zoom during a period of coherent oscillations. Bottom: Power spectra of the mass-weighted radius computed over one tearing cycle (grey curve) and over a period of time in which the disk does not tear (dashed green curve). The vertical dashed black line shows the radial epicyclic frequency at $13 r_G$ (55 Hz for a $10 M_\odot$ black hole). Image reproduced by permission from \cite{Musoke23}, copyright by RAS.}
\label{fig:Musoke}       
\end{figure}

It now appears that it may be necessary to have both tilt {\em and} a distinct transition radius in the disk to trigger sustained QPO activity \cite{Bollimpalli24}. For example, \cite{Musoke23} found that, in a tearing disk simulation (tearing disks were discussed in Section \ref{sec:tearing}), the inner accretion disk oscillates at the {\em radial} epicyclic frequency associated with the tearing radius (see Figure \ref{fig:Musoke}). Oscillations at {\em local} mode frequencies have been seen in many accretion disk simulations \cite{Reynolds09, Mishra19}, but because each radius excites a different frequency, these usually do not produce a QPO-like peak. The key in the tearing disk results is that the tearing radius provides a preferred frequency that is amplified with respect to the others. 

\begin{figure}
\sidecaption
\includegraphics[width=1.0\linewidth,trim=0mm 0mm 0mm 0,clip]{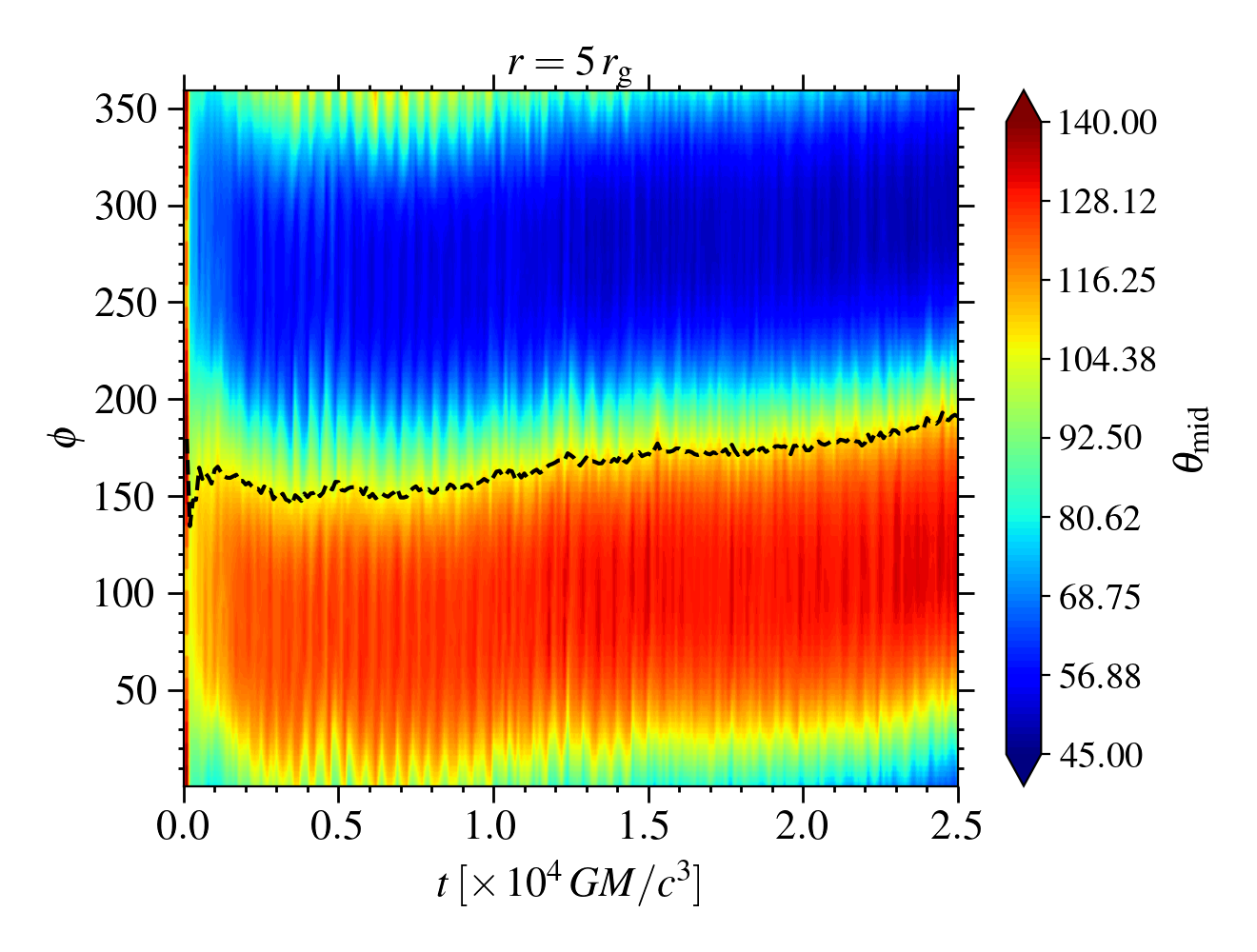}
%
%
\caption{A time–space plot of the mid-plane angle, $\theta_\mathrm{mid}$, near the inner edge of a tilted, truncated disk simulation. A clear $m = 1$ pattern is seen, which precesses along with the rest of the thick-disk region (black, dashed line) and also oscillates with a higher-frequency period close to $500 GM/c^3$. Image reproduced by permission from \cite{Bollimpalli24}, copyright by RAS.}
\label{fig:Deepika}       
\end{figure}

However, it may not be enough just to have a transition radius; tilt also appears to be important. For instance, \cite{Bollimpalli24} found excitation of {\em vertical} epicyclic motion in their truncated disk simulations, but only ones that included tilt. Figure \ref{fig:Deepika} provides visual evidence for both low- and high-frequency variations, specifically in the midplane $\theta$ value, for one such tilted, truncated disk simulation.

\section{Future Directions}
\label{sec:future}

In this chapter, we have reviewed much of the interesting phenomenology associated with simulations of tilted accretion disks, including precession (Section \ref{sec:precession}), the BP effect (Section \ref{sec:BPeffect}), and disk tearing (Section \ref{sec:tearing}). While we gave some basic arguments for when the disk might undergo each of these processes (see Sections \ref{sec:overview} and \ref{sec:tearing}), there are still large uncertainties within the available parameter space and many more simulations need to be done to fill in the gaps. In this closing section, we outline what we see as some of the most promising directions to explore in the near future.

One very important issue to settle is when do the jets associated with tilted disks precess and when do they align? As we mentioned in Section \ref{sec:jets}, jets in some simulations appear to precess with the same frequency as the inner disk, while in others, specifically tilted, MAD simulations, the whole inner disk is forced to align with the black hole and the jet does not precess. However, these conclusions are currently based on a very small number of simulations. More work needs to be done to flesh out the rough conclusion that SANE disks and jets precess, while MAD ones do not. This is pertinent to, among other things, the question of whether the hard state of BHBs can be MAD \cite{Fragile23}. There is also a question as to what is the distance scale over which jet precession can be maintained, and therefore, observed? This is critically important as it may be through X-ray or radio jets that LT precession is first confirmed in nature (see e.g. \cite{Miller-Jones19}).

Another issue is that current GRMHD simulations \cite{Liska19b, Liska20} suggest that, for tearing to occur, the disk needs be extremely thin ($\delta \lesssim 0.02 $) and highly misaligned ($\beta \gtrsim 45^{\circ}$). This would imply that tearing can only occur at luminosities $L \gtrsim 0.01 L_{\rm Edd}$ and for highly misaligned sources, which would seem to make it inapplicable to most BHBs. However, tearing from a thin disk might not be the only avenue to form a precessing disk. If some other process (e.g., thermal conduction and evaporation) can separate the disk into distinct components, then it may be the inner one can precess semi-independent of the outer. Or, super-Eddington disks might precess inside their spherization radius \cite{Middleton18}. Finally, GRMHD simulations have yet to consider even thinner disks ($\delta \lesssim 0.02$) and whether they can tear at smaller misalignment angles.

Speaking of super-Eddington accretion, most ultra-luminous X-ray sources (ULXs) are now believed to be powered by stellar-mass compact objects accreting at super-Eddington rates\footnote{See \cite{Kaaret17, King23} for reviews of ULXs.}. What effect would tilt have in those cases? How does the addition of dynamically important radiative forces within the disk interplay with the LT torques to change the resulting disk structure? How apparent would those changes be out at the effective photosphere of the disk, i.e., is there an observational signature of tilted ULXs? Here, LT warping may also compete with radiative warping \cite{Pringle96}, such that the final disk structure could be quite complex. Currently, there is very little published work in this regime. 

What about LT precession in TDEs? As we argued in Section \ref{sec:ubiquity}, TDE disks should almost always be tilted. Is there some way to exploit TDE light curves to look for evidence of LT precession or BP alignment? Before we can answer that question, we probably need more simulations of TDEs that properly account for the effects of black hole spin (i.e., general relativistic hydro, MHD, or even radiation MHD simulations) and consider a range of inclinations for the encounter. Interestingly, some TDEs show QPO signals that may be consistent with LT precession \cite{Pasham19, Pasham24}, so there is motivation to do this work.

Finally, Sgr A*, the $4\times10^6 M_\odot$ black hole at the center of the Milky Way, is another important source to compare simulations with. In this case, the evidence points to the black hole primarily being fed by winds from Wolf–Rayet stars orbiting near the Galactic center, making it possible to model the entire accretion flow from source to black hole \cite{Ressler18, Ressler23}. Because these stars orbit on scales that are unlikely to be affected by the spin axis of Sgr A*, this again likely means the accretion flow is tilted and might even be warped or torn. Depending on the scale of the warping, instruments such as the Event Horizon Telescope \cite{EHT22} or GRAVITY \cite{GRAVITY18} may even give us a chance to directly image a tilted, warped accretion disk. On that hopeful note, we end our chapter.


\begin{acknowledgement}
PCF gratefully acknowledges the support of NASA through award No 23-ATP23-0100.
\end{acknowledgement}



\begin{thebibliography}{10}

\bibitem{Bambic24}
Christopher~J. {Bambic}, Eliot {Quataert}, Matthew~W. {Kunz}, and Yan-Fei
  {Jiang}.
\newblock {Local models of two-temperature accretion disc coronae. II. Ion
  thermal conduction and the absence of disc evaporation}.
\newblock {\em arXiv e-prints}, page arXiv:2401.05482, January 2024.

\bibitem{Bardeen75}
James~M. {Bardeen} and Jacobus~A. {Petterson}.
\newblock {The Lense-Thirring Effect and Accretion Disks around Kerr Black
  Holes}.
\newblock {\em \apjl}, 195:L65, January 1975.

\bibitem{Blandford82}
R.~D. {Blandford} and D.~G. {Payne}.
\newblock {Hydromagnetic flows from accretion discs and the production of radio
  jets}.
\newblock {\em \mnras}, 199:883--903, June 1982.

\bibitem{Blandford77}
R.~D. {Blandford} and R.~L. {Znajek}.
\newblock {Electromagnetic extraction of energy from Kerr black holes}.
\newblock {\em \mnras}, 179:433--456, May 1977.

\bibitem{Bollimpalli24}
D.~A. {Bollimpalli}, P.~C. {Fragile}, J.~W. {Dewberry}, and W.~{Klu{\'z}niak}.
\newblock {Truncated, tilted discs as a possible source of Quasi-Periodic
  Oscillations}.
\newblock {\em \mnras}, 528(2):1142--1157, February 2024.

\bibitem{Bollimpalli23}
D.~A. {Bollimpalli}, P.~C. {Fragile}, and W.~{Klu{\'z}niak}.
\newblock {Effect of geometrically thin discs on precessing, thick flows:
  relevance to type-C QPOs}.
\newblock {\em \mnras}, 520(1):L79--L84, March 2023.

\bibitem{Caproni07}
A.~{Caproni}, Z.~{Abraham}, M.~{Livio}, and H.~J. {Mosquera Cuesta}.
\newblock {Is the Bardeen-Petterson effect responsible for the warping and
  precession in NGC4258?}
\newblock {\em \mnras}, 379(1):135--142, July 2007.

\bibitem{Caproni06}
Anderson {Caproni}, Zulema {Abraham}, and Herman~J. {Mosquera Cuesta}.
\newblock {Bardeen-Petterson Effect and the Disk Structure of the Seyfert
  Galaxy NGC 1068}.
\newblock {\em \apj}, 638(1):120--124, February 2006.

\bibitem{Chatterjee19}
K.~{Chatterjee}, M.~{Liska}, A.~{Tchekhovskoy}, and S.~B. {Markoff}.
\newblock {Accelerating AGN jets to parsec scales using general relativistic
  MHD simulations}.
\newblock {\em \mnras}, 490(2):2200--2218, Dec 2019.

\bibitem{Chatterjee23}
Koushik {Chatterjee}, Matthew {Liska}, Alexander {Tchekhovskoy}, and Sera
  {Markoff}.
\newblock {Misaligned magnetized accretion flows onto spinning black holes:
  magneto-spin alignment, outflow power and intermittent jets}.
\newblock {\em arXiv e-prints}, page arXiv:2311.00432, November 2023.

\bibitem{Cheng16}
Yifan {Cheng}, Dan {Liu}, Sourabh {Nampalliwar}, and Cosimo {Bambi}.
\newblock {X-ray spectropolarimetric signature of a warped disk around a
  stellar-mass black hole}.
\newblock {\em Classical and Quantum Gravity}, 33(12):125015, June 2016.

\bibitem{Cho22}
Hyerin {Cho} and Ramesh {Narayan}.
\newblock {Analytical Model of Disk Evaporation and State Transitions in
  Accreting Black Holes}.
\newblock {\em \apj}, 932(2):97, June 2022.

\bibitem{Dexter11}
Jason {Dexter} and P.~Chris {Fragile}.
\newblock {Observational Signatures of Tilted Black Hole Accretion Disks from
  Simulations}.
\newblock {\em \apj}, 730(1):36, March 2011.

\bibitem{Done07}
C.~{Done}, M.~{Gierli{\'n}ski}, and A.~{Kubota}.
\newblock {Modelling the behaviour of accretion flows in X-ray binaries.
  Everything you always wanted to know about accretion but were afraid to ask}.
\newblock {\em \aapr}, 15:1--66, December 2007.

\bibitem{Dogan18}
S.~{Do{\v g}an}, C.~J. {Nixon}, A.~R. {King}, and J.~E. {Pringle}.
\newblock {Instability of warped discs}.
\newblock {\em \mnras}, 476:1519--1531, May 2018.

\bibitem{Esin97}
Ann~A. {Esin}, Jeffrey~E. {McClintock}, and Ramesh {Narayan}.
\newblock {Advection-Dominated Accretion and the Spectral States of Black Hole
  X-Ray Binaries: Application to Nova Muscae 1991}.
\newblock {\em \apj}, 489(2):865--889, November 1997.

\bibitem{EHT22}
{Event Horizon Telescope Collaboration}, Kazunori {Akiyama}, Antxon {Alberdi},
  Walter {Alef}, Juan~Carlos {Algaba}, Richard {Anantua}, Keiichi {Asada},
  Rebecca {Azulay}, Uwe {Bach}, Anne-Kathrin {Baczko}, David {Ball}, Mislav
  {Balokovi{\'c}}, John {Barrett}, Michi {Baub{\"o}ck}, Bradford~A. {Benson},
  Dan {Bintley}, Lindy {Blackburn}, Raymond {Blundell}, Katherine~L. {Bouman},
  Geoffrey~C. {Bower}, Hope {Boyce}, Michael {Bremer}, Christiaan~D.
  {Brinkerink}, Roger {Brissenden}, Silke {Britzen}, Avery~E. {Broderick},
  Dominique {Broguiere}, Thomas {Bronzwaer}, Sandra {Bustamante}, Do-Young
  {Byun}, John~E. {Carlstrom}, Chiara {Ceccobello}, Andrew {Chael}, Chi-kwan
  {Chan}, Koushik {Chatterjee}, Shami {Chatterjee}, Ming-Tang {Chen}, Yongjun
  {Chen}, Xiaopeng {Cheng}, Ilje {Cho}, Pierre {Christian}, Nicholas~S.
  {Conroy}, John~E. {Conway}, James~M. {Cordes}, Thomas~M. {Crawford},
  Geoffrey~B. {Crew}, Alejandro {Cruz-Osorio}, Yuzhu {Cui}, Jordy {Davelaar},
  Mariafelicia {De Laurentis}, Roger {Deane}, Jessica {Dempsey}, Gregory
  {Desvignes}, Jason {Dexter}, Vedant {Dhruv}, Sheperd~S. {Doeleman}, Sean
  {Dougal}, Sergio~A. {Dzib}, Ralph~P. {Eatough}, Razieh {Emami}, Heino
  {Falcke}, Joseph {Farah}, Vincent~L. {Fish}, Ed~{Fomalont}, H.~Alyson {Ford},
  Raquel {Fraga-Encinas}, William~T. {Freeman}, Per {Friberg}, Christian~M.
  {Fromm}, Antonio {Fuentes}, Peter {Galison}, Charles~F. {Gammie}, Roberto
  {Garc{\'\i}a}, Olivier {Gentaz}, Boris {Georgiev}, Ciriaco {Goddi}, Roman
  {Gold}, Arturo~I. {G{\'o}mez-Ruiz}, Jos{\'e}~L. {G{\'o}mez}, Minfeng {Gu},
  Mark {Gurwell}, Kazuhiro {Hada}, Daryl {Haggard}, Kari {Haworth}, Michael~H.
  {Hecht}, Ronald {Hesper}, Dirk {Heumann}, Luis~C. {Ho}, Paul {Ho}, Mareki
  {Honma}, Chih-Wei~L. {Huang}, Lei {Huang}, David~H. {Hughes}, Shiro {Ikeda},
  C.~M.~Violette {Impellizzeri}, Makoto {Inoue}, Sara {Issaoun}, David~J.
  {James}, Buell~T. {Jannuzi}, Michael {Janssen}, Britton {Jeter}, Wu~{Jiang},
  Alejandra {Jim{\'e}nez-Rosales}, Michael~D. {Johnson}, Svetlana {Jorstad},
  Abhishek~V. {Joshi}, Taehyun {Jung}, Mansour {Karami}, Ramesh {Karuppusamy},
  Tomohisa {Kawashima}, Garrett~K. {Keating}, Mark {Kettenis}, Dong-Jin {Kim},
  Jae-Young {Kim}, Jongsoo {Kim}, Junhan {Kim}, Motoki {Kino}, Jun~Yi {Koay},
  Prashant {Kocherlakota}, Yutaro {Kofuji}, Patrick~M. {Koch}, Shoko {Koyama},
  Carsten {Kramer}, Michael {Kramer}, Thomas~P. {Krichbaum}, Cheng-Yu {Kuo},
  Noemi {La Bella}, Tod~R. {Lauer}, Daeyoung {Lee}, Sang-Sung {Lee}, Po~Kin
  {Leung}, Aviad {Levis}, Zhiyuan {Li}, Rocco {Lico}, Greg {Lindahl}, Michael
  {Lindqvist}, Mikhail {Lisakov}, Jun {Liu}, Kuo {Liu}, Elisabetta {Liuzzo},
  Wen-Ping {Lo}, Andrei~P. {Lobanov}, Laurent {Loinard}, Colin~J. {Lonsdale},
  Ru-Sen {Lu}, Jirong {Mao}, Nicola {Marchili}, Sera {Markoff}, Daniel~P.
  {Marrone}, Alan~P. {Marscher}, Iv{\'a}n {Mart{\'\i}-Vidal}, Satoki
  {Matsushita}, Lynn~D. {Matthews}, Lia {Medeiros}, Karl~M. {Menten}, Daniel
  {Michalik}, Izumi {Mizuno}, Yosuke {Mizuno}, James~M. {Moran}, Kotaro
  {Moriyama}, Monika {Moscibrodzka}, Cornelia {M{\"u}ller}, Alejandro {Mus},
  Gibwa {Musoke}, Ioannis {Myserlis}, Andrew {Nadolski}, Hiroshi {Nagai},
  Neil~M. {Nagar}, Masanori {Nakamura}, Ramesh {Narayan}, Gopal {Narayanan},
  Iniyan {Natarajan}, Antonios {Nathanail}, Santiago~Navarro {Fuentes}, Joey
  {Neilsen}, Roberto {Neri}, Chunchong {Ni}, Aristeidis {Noutsos}, Michael~A.
  {Nowak}, Junghwan {Oh}, Hiroki {Okino}, H{\'e}ctor {Olivares}, Gisela~N.
  {Ortiz-Le{\'o}n}, Tomoaki {Oyama}, Feryal {{\"O}zel}, Daniel C.~M. {Palumbo},
  Georgios~Filippos {Paraschos}, Jongho {Park}, Harriet {Parsons}, Nimesh
  {Patel}, Ue-Li {Pen}, Dominic~W. {Pesce}, Vincent {Pi{\'e}tu}, Richard
  {Plambeck}, Aleksandar {PopStefanija}, Oliver {Porth}, Felix~M. {P{\"o}tzl},
  Ben {Prather}, Jorge~A. {Preciado-L{\'o}pez}, Dimitrios {Psaltis}, Hung-Yi
  {Pu}, Venkatessh {Ramakrishnan}, Ramprasad {Rao}, Mark~G. {Rawlings},
  Alexander~W. {Raymond}, Luciano {Rezzolla}, Angelo {Ricarte}, Bart
  {Ripperda}, Freek {Roelofs}, Alan {Rogers}, Eduardo {Ros}, Cristina
  {Romero-Ca{\~n}izales}, Arash {Roshanineshat}, Helge {Rottmann}, Alan~L.
  {Roy}, Ignacio {Ruiz}, Chet {Ruszczyk}, Kazi L.~J. {Rygl}, Salvador
  {S{\'a}nchez}, David {S{\'a}nchez-Arg{\"u}elles}, Miguel
  {S{\'a}nchez-Portal}, Mahito {Sasada}, Kaushik {Satapathy}, Tuomas
  {Savolainen}, F.~Peter {Schloerb}, Jonathan {Schonfeld}, Karl-Friedrich
  {Schuster}, Lijing {Shao}, Zhiqiang {Shen}, Des {Small}, Bong~Won {Sohn},
  Jason {SooHoo}, Kamal {Souccar}, He~{Sun}, Fumie {Tazaki}, Alexandra~J.
  {Tetarenko}, Paul {Tiede}, Remo P.~J. {Tilanus}, Michael {Titus}, Pablo
  {Torne}, Efthalia {Traianou}, Tyler {Trent}, Sascha {Trippe}, Matthew {Turk},
  Ilse {van Bemmel}, Huib~Jan {van Langevelde}, Daniel~R. {van Rossum}, Jesse
  {Vos}, Jan {Wagner}, Derek {Ward-Thompson}, John {Wardle}, Jonathan
  {Weintroub}, Norbert {Wex}, Robert {Wharton}, Maciek {Wielgus}, Kaj {Wiik},
  Gunther {Witzel}, Michael~F. {Wondrak}, George~N. {Wong}, Qingwen {Wu}, Paul
  {Yamaguchi}, Doosoo {Yoon}, Andr{\'e} {Young}, Ken {Young}, Ziri {Younsi},
  Feng {Yuan}, Ye-Fei {Yuan}, J.~Anton {Zensus}, Shuo {Zhang}, Guang-Yao
  {Zhao}, Shan-Shan {Zhao}, Claudio {Agurto}, Alexander {Allardi}, Rodrigo
  {Amestica}, Juan~Pablo {Araneda}, Oriel {Arriagada}, Jennie~L. {Berghuis},
  Alessandra {Bertarini}, Ryan {Berthold}, Jay {Blanchard}, Ken {Brown},
  Mauricio {C{\'a}rdenas}, Michael {Cantzler}, Patricio {Caro}, Edgar
  {Castillo-Dom{\'\i}nguez}, Tin~Lok {Chan}, Chih-Cheng {Chang}, Dominic~O.
  {Chang}, Shu-Hao {Chang}, Song-Chu {Chang}, Chung-Chen {Chen}, Ryan
  {Chilson}, Tim~C. {Chuter}, Miroslaw {Ciechanowicz}, Edgar {Colin-Beltran},
  Iain~M. {Coulson}, Joseph {Crowley}, Nathalie {Degenaar}, Sven {Dornbusch},
  Carlos~A. {Dur{\'a}n}, Wendeline~B. {Everett}, Aaron {Faber}, Karl {Forster},
  Miriam~M. {Fuchs}, David~M. {Gale}, Gertie {Geertsema}, Edouard
  {Gonz{\'a}lez}, Dave {Graham}, Fr{\'e}d{\'e}ric {Gueth}, Nils~W. {Halverson},
  Chih-Chiang {Han}, Kuo-Chang {Han}, Yutaka {Hasegawa}, Jos{\'e}~Luis
  {Hern{\'a}ndez-Rebollar}, Cristian {Herrera}, Ruben {Herrero-Illana}, Stefan
  {Heyminck}, Akihiko {Hirota}, James {Hoge}, Shelbi~R. {Hostler Schimpf},
  Ryan~E. {Howie}, Yau-De {Huang}, Homin {Jiang}, Hao {Jinchi}, David {John},
  Kimihiro {Kimura}, Thomas {Klein}, Derek {Kubo}, John {Kuroda}, Caleb {Kwon},
  Richard {Lacasse}, Robert {Laing}, Erik~M. {Leitch}, Chao-Te {Li}, Ching-Tang
  {Liu}, Kuan-Yu {Liu}, Lupin C.~C. {Lin}, Li-Ming {Lu}, Felipe {Mac-Auliffe},
  Pierre {Martin-Cocher}, Callie {Matulonis}, John~K. {Maute}, Hugo {Messias},
  Zheng {Meyer-Zhao}, Alfredo {Monta{\~n}a}, Francisco {Montenegro-Montes},
  William {Montgomerie}, Marcos~Emir {Moreno Nolasco}, Dirk {Muders}, Hiroaki
  {Nishioka}, Timothy~J. {Norton}, George {Nystrom}, Hideo {Ogawa}, Rodrigo
  {Olivares}, Peter {Oshiro}, Juan~Pablo {P{\'e}rez-Beaupuits}, Rodrigo
  {Parra}, Neil~M. {Phillips}, Michael {Poirier}, Nicolas {Pradel}, Richard
  {Qiu}, Philippe~A. {Raffin}, Alexandra~S. {Rahlin}, Jorge {Ram{\'\i}rez},
  Sean {Ressler}, Mark {Reynolds}, Iv{\'a}n {Rodr{\'\i}guez-Montoya},
  Alejandro~F. {Saez-Madain}, Jorge {Santana}, Paul {Shaw}, Leslie~E.
  {Shirkey}, Kevin~M. {Silva}, William {Snow}, Don {Sousa}, T.~K. {Sridharan},
  William {Stahm}, Anthony~A. {Stark}, John {Test}, Karl {Torstensson}, Paulina
  {Venegas}, Craig {Walther}, Ta-Shun {Wei}, Chris {White}, Gundolf {Wieching},
  Rudy {Wijnands}, Jan G.~A. {Wouterloot}, Chen-Yu {Yu}, Wei {Yu}, and Milagros
  {Zeballos}.
\newblock {First Sagittarius A* Event Horizon Telescope Results. I. The Shadow
  of the Supermassive Black Hole in the Center of the Milky Way}.
\newblock {\em \apjl}, 930(2):L12, May 2022.

\bibitem{Ferreira06}
J.~{Ferreira}, P.~O. {Petrucci}, G.~{Henri}, L.~{Saug{\'e}}, and
  G.~{Pelletier}.
\newblock {A unified accretion-ejection paradigm for black hole X-ray binaries.
  I. The dynamical constituents}.
\newblock {\em \aap}, 447:813--825, March 2006.

\bibitem{Fragile09b}
P.~Chris {Fragile}.
\newblock {Effective Inner Radius of Tilted Black Hole Accretion Disks}.
\newblock {\em \apjl}, 706(2):L246--L250, December 2009.

\bibitem{Fragile05}
P.~Chris {Fragile} and Peter {Anninos}.
\newblock {Hydrodynamic Simulations of Tilted Thick-Disk Accretion onto a Kerr
  Black Hole}.
\newblock {\em \apj}, 623(1):347--361, April 2005.

\bibitem{Fragile08}
P.~Chris {Fragile} and Omer~M. {Blaes}.
\newblock {Epicyclic Motions and Standing Shocks in Numerically Simulated
  Tilted Black Hole Accretion Disks}.
\newblock {\em \apj}, 687(2):757--766, November 2008.

\bibitem{Fragile07}
P.~Chris {Fragile}, Omer~M. {Blaes}, Peter {Anninos}, and Jay~D. {Salmonson}.
\newblock {Global General Relativistic Magnetohydrodynamic Simulation of a
  Tilted Black Hole Accretion Disk}.
\newblock {\em \apj}, 668(1):417--429, October 2007.

\bibitem{Fragile23}
P.~Chris {Fragile}, Koushik {Chatterjee}, Adam {Ingram}, and Matthew
  {Middleton}.
\newblock {The luminous, hard state can't be MAD}.
\newblock {\em \mnras}, 525(1):L82--L86, October 2023.

\bibitem{Fragile09}
P.~Chris {Fragile}, Christopher~C. {Lindner}, Peter {Anninos}, and Jay~D.
  {Salmonson}.
\newblock {Application of the Cubed-Sphere Grid to Tilted Black Hole Accretion
  Disks}.
\newblock {\em \apj}, 691(1):482--494, January 2009.

\bibitem{Fragile01}
P.~Chris {Fragile}, Grant~J. {Mathews}, and James~R. {Wilson}.
\newblock {Bardeen-Petterson Effect and Quasi-periodic Oscillations in X-Ray
  Binaries}.
\newblock {\em \apj}, 553(2):955--959, June 2001.

\bibitem{Fragos10}
T.~{Fragos}, M.~{Tremmel}, E.~{Rantsiou}, and K.~{Belczynski}.
\newblock {Black Hole Spin-Orbit Misalignment in Galactic X-ray Binaries}.
\newblock {\em \apjl}, 719(1):L79--L83, August 2010.

\bibitem{Generozov14}
Aleksey {Generozov}, Omer {Blaes}, P.~Chris {Fragile}, and Ken~B. {Henisey}.
\newblock {Physical Properties of the Inner Shocks in Hot, Tilted Black Hole
  Accretion Flows}.
\newblock {\em \apj}, 780(1):81, January 2014.

\bibitem{GRAVITY18}
{GRAVITY Collaboration}, R.~{Abuter}, A.~{Amorim}, M.~{Baub{\"o}ck}, J.~P.
  {Berger}, H.~{Bonnet}, W.~{Brandner}, Y.~{Cl{\'e}net}, V.~{Coud{\'e} Du
  Foresto}, P.~T. {de Zeeuw}, C.~{Deen}, J.~{Dexter}, G.~{Duvert}, A.~{Eckart},
  F.~{Eisenhauer}, N.~M. {F{\"o}rster Schreiber}, P.~{Garcia}, F.~{Gao},
  E.~{Gendron}, R.~{Genzel}, S.~{Gillessen}, P.~{Guajardo}, M.~{Habibi},
  X.~{Haubois}, Th. {Henning}, S.~{Hippler}, M.~{Horrobin}, A.~{Huber},
  A.~{Jim{\'e}nez-Rosales}, L.~{Jocou}, P.~{Kervella}, S.~{Lacour},
  V.~{Lapeyr{\`e}re}, B.~{Lazareff}, J.~B. {Le Bouquin}, P.~{L{\'e}na},
  M.~{Lippa}, T.~{Ott}, J.~{Panduro}, T.~{Paumard}, K.~{Perraut}, G.~{Perrin},
  O.~{Pfuhl}, P.~M. {Plewa}, S.~{Rabien}, G.~{Rodr{\'\i}guez-Coira},
  G.~{Rousset}, A.~{Sternberg}, O.~{Straub}, C.~{Straubmeier}, E.~{Sturm},
  L.~J. {Tacconi}, F.~{Vincent}, S.~{von Fellenberg}, I.~{Waisberg},
  F.~{Widmann}, E.~{Wieprecht}, E.~{Wiezorrek}, J.~{Woillez}, and S.~{Yazici}.
\newblock {Detection of orbital motions near the last stable circular orbit of
  the massive black hole SgrA*}.
\newblock {\em \aap}, 618:L10, October 2018.

\bibitem{Hannikainen01}
Diana {Hannikainen}, Duncan {Campbell-Wilson}, Richard {Hunstead}, Vince
  {McIntyre}, Jim {Lovell}, John {Reynolds}, Tasso {Tzioumis}, and Kinwah {Wu}.
\newblock {XTE J1550-564: a superluminal ejection during the September 1998
  outburst}.
\newblock {\em Astrophysics and Space Science Supplement}, 276:45--48, January
  2001.

\bibitem{Heil15}
L.~M. {Heil}, P.~{Uttley}, and M.~{Klein-Wolt}.
\newblock {Inclination-dependent spectral and timing properties in transient
  black hole X-ray binaries}.
\newblock {\em \mnras}, 448(4):3348--3353, April 2015.

\bibitem{Henisey12}
Ken~B. {Henisey}, Omer~M. {Blaes}, and P.~Chris {Fragile}.
\newblock {Variability from Non-axisymmetric Fluctuations Interacting with
  Standing Shocks in Tilted Black Hole Accretion Disks}.
\newblock {\em \apj}, 761(1):18, December 2012.

\bibitem{Henisey09}
Ken~B. {Henisey}, Omer~M. {Blaes}, P.~Chris {Fragile}, and B{\'a}rbara~T.
  {Ferreira}.
\newblock {Excitation of Trapped Waves in Simulations of Tilted Black Hole
  Accretion Disks with Magnetorotational Turbulence}.
\newblock {\em \apj}, 706(1):705--711, November 2009.

\bibitem{Ingram11}
Adam {Ingram} and Chris {Done}.
\newblock {A physical model for the continuum variability and quasi-periodic
  oscillation in accreting black holes}.
\newblock {\em \mnras}, 415(3):2323--2335, August 2011.

\bibitem{Ingram09}
Adam {Ingram}, Chris {Done}, and P.~Chris {Fragile}.
\newblock {Low-frequency quasi-periodic oscillations spectra and Lense-Thirring
  precession}.
\newblock {\em \mnras}, 397(1):L101--L105, July 2009.

\bibitem{Ingram16}
Adam {Ingram}, Michiel {van der Klis}, Matthew {Middleton}, Chris {Done}, Diego
  {Altamirano}, Lucy {Heil}, Phil {Uttley}, and Magnus {Axelsson}.
\newblock {A quasi-periodic modulation of the iron line centroid energy in the
  black hole binary H1743-322}.
\newblock {\em \mnras}, 461(2):1967--1980, September 2016.

\bibitem{Ingram19}
Adam~R. {Ingram} and Sara~E. {Motta}.
\newblock {A review of quasi-periodic oscillations from black hole X-ray
  binaries: Observation and theory}.
\newblock {\em \nar}, 85:101524, September 2019.

\bibitem{Ivanov97}
P.~B. {Ivanov} and A.~F. {Illarionov}.
\newblock {The oscillatory shape of the stationary twisted disc around a Kerr
  black hole}.
\newblock {\em \mnras}, 285(2):394--402, February 1997.

\bibitem{Kaaret17}
Philip {Kaaret}, Hua {Feng}, and Timothy~P. {Roberts}.
\newblock {Ultraluminous X-Ray Sources}.
\newblock {\em \araa}, 55(1):303--341, August 2017.

\bibitem{Kaaz22}
Nicholas {Kaaz}, Matthew T.~P. {Liska}, Jonatan {Jacquemin-Ide}, Zachary~L.
  {Andalman}, Gibwa {Musoke}, Alexander {Tchekhovskoy}, and Oliver {Porth}.
\newblock {Nozzle Shocks, Disk Tearing and Streamers Drive Rapid Accretion in
  3D GRMHD Simulations of Warped Thin Disks}.
\newblock {\em arXiv e-prints}, page arXiv:2210.10053, October 2022.

\bibitem{King08}
A.~R. {King}, J.~E. {Pringle}, and J.~A. {Hofmann}.
\newblock {The evolution of black hole mass and spin in active galactic
  nuclei}.
\newblock {\em \mnras}, 385(3):1621--1627, April 2008.

\bibitem{King23}
Andrew {King}, Jean-Pierre {Lasota}, and Matthew {Middleton}.
\newblock {Ultraluminous X-ray sources}.
\newblock {\em \nar}, 96:101672, June 2023.

\bibitem{Kinney00}
A.~L. {Kinney}, H.~R. {Schmitt}, C.~J. {Clarke}, J.~E. {Pringle}, J.~S.
  {Ulvestad}, and R.~R.~J. {Antonucci}.
\newblock {Jet Directions in Seyfert Galaxies}.
\newblock {\em \apj}, 537(1):152--177, July 2000.

\bibitem{Kondratko05}
Paul~T. {Kondratko}, Lincoln~J. {Greenhill}, and James~M. {Moran}.
\newblock {Evidence for a Geometrically Thick Self-Gravitating Accretion Disk
  in NGC 3079}.
\newblock {\em \apj}, 618(2):618--634, January 2005.

\bibitem{Kumar85}
S.~{Kumar} and J.~E. {Pringle}.
\newblock {Twisted accretion discs - The Bardeen-Petterson effect}.
\newblock {\em \mnras}, 213:435--442, March 1985.

\bibitem{Larwood96}
J.~D. {Larwood}, R.~P. {Nelson}, J.~C.~B. {Papaloizou}, and C.~{Terquem}.
\newblock {The tidally induced warping, precession and truncation of accretion
  discs in binary systems: three-dimensional simulations}.
\newblock {\em \mnras}, 282(2):597--613, September 1996.

\bibitem{Lense18}
Josef {Lense} and Hans {Thirring}.
\newblock {{\"U}ber den Einflu{\ss} der Eigenrotation der Zentralk{\"o}rper auf
  die Bewegung der Planeten und Monde nach der Einsteinschen
  Gravitationstheorie}.
\newblock {\em Physikalische Zeitschrift}, 19:156, January 1918.

\bibitem{Liska18}
M.~{Liska}, C.~{Hesp}, A.~{Tchekhovskoy}, A.~{Ingram}, M.~{van der Klis}, and
  S.~{Markoff}.
\newblock {Formation of precessing jets by tilted black hole discs in 3D
  general relativistic MHD simulations}.
\newblock {\em \mnras}, 474(1):L81--L85, February 2018.

\bibitem{Liska23}
M.~{Liska}, C.~{Hesp}, A.~{Tchekhovskoy}, A.~{Ingram}, M.~{van der Klis}, and
  S.~B. {Markoff}.
\newblock {A phase lag between disc and corona in GRMHD simulations of
  precessing tilted accretion discs}.
\newblock {\em \na}, 101:102012, July 2023.

\bibitem{Liska19b}
M.~{Liska}, C.~{Hesp}, A.~{Tchekhovskoy}, A.~{Ingram}, M.~{van der Klis}, S.~B.
  {Markoff}, and M.~{Van Moer}.
\newblock {Disc Tearing and Bardeen-Petterson Alignment in GRMHD Simulations of
  Highly Tilted Thin Accretion Discs}.
\newblock {\em arXiv e-prints}, page arXiv:1904.08428, Apr 2019.

\bibitem{Liska19}
M.~{Liska}, A.~{Tchekhovskoy}, A.~{Ingram}, and M.~{van der Klis}.
\newblock {Bardeen-Petterson alignment, jets, and magnetic truncation in GRMHD
  simulations of tilted thin accretion discs}.
\newblock {\em \mnras}, 487(1):550--561, July 2019.

\bibitem{Liska20}
M.~T.~P. {Liska}, K.~{Chatterjee}, D.~{Issa}, D.~{Yoon}, N.~{Kaaz},
  A.~{Tchekhovskoy}, D.~{van Eijnatten}, G.~{Musoke}, C.~{Hesp}, V.~{Rohoza},
  S.~{Markoff}, A.~{Ingram}, and M.~{van der Klis}.
\newblock {H-AMR: A New GPU-accelerated GRMHD Code for Exascale Computing with
  3D Adaptive Mesh Refinement and Local Adaptive Time Stepping}.
\newblock {\em \apjs}, 263(2):26, December 2022.

\bibitem{Liska23b}
M.~T.~P. {Liska}, N.~{Kaaz}, G.~{Musoke}, A.~{Tchekhovskoy}, and O.~{Porth}.
\newblock {Radiation Transport Two-temperature GRMHD Simulations of Warped
  Accretion Disks}.
\newblock {\em \apjl}, 944(2):L48, February 2023.

\bibitem{Liska22}
M.~T.~P. {Liska}, G.~{Musoke}, A.~{Tchekhovskoy}, O.~{Porth}, and A.~M.
  {Beloborodov}.
\newblock {Formation of Magnetically Truncated Accretion Disks in 3D
  Radiation-transport Two-temperature GRMHD Simulations}.
\newblock {\em \apjl}, 935(1):L1, August 2022.

\bibitem{Liu02}
Siming {Liu} and Fulvio {Melia}.
\newblock {Spin-induced Disk Precession in the Supermassive Black Hole at the
  Galactic Center}.
\newblock {\em \apjl}, 573(1):L23--L26, July 2002.

\bibitem{Lodato10}
Giuseppe {Lodato} and Daniel~J. {Price}.
\newblock {On the diffusive propagation of warps in thin accretion discs}.
\newblock {\em \mnras}, 405(2):1212--1226, June 2010.

\bibitem{Lodato07}
Giuseppe {Lodato} and J.~E. {Pringle}.
\newblock {Warp diffusion in accretion discs: a numerical investigation}.
\newblock {\em \mnras}, 381(3):1287--1300, November 2007.

\bibitem{Maccarone02}
Thomas~J. {Maccarone}.
\newblock {On the misalignment of jets in microquasars}.
\newblock {\em \mnras}, 336(4):1371--1376, November 2002.

\bibitem{Marcel2020}
G.~{Marcel}, F.~{Cangemi}, J.~{Rodriguez}, J.~{Neilsen}, J.~{Ferreira}, P.~O.
  {Petrucci}, J.~{Malzac}, S.~{Barnier}, and M.~{Clavel}.
\newblock {A unified accretion-ejection paradigm for black hole X-ray binaries.
  V. Low-frequency quasi-periodic oscillations}.
\newblock {\em \aap}, 640:A18, August 2020.

\bibitem{McKinney06}
Jonathan~C. {McKinney}.
\newblock {General relativistic magnetohydrodynamic simulations of the jet
  formation and large-scale propagation from black hole accretion systems}.
\newblock {\em \mnras}, 368(4):1561--1582, June 2006.

\bibitem{McKinney13}
Jonathan~C. {McKinney}, Alexander {Tchekhovskoy}, and Roger~D. {Blandford}.
\newblock {Alignment of Magnetized Accretion Disks and Relativistic Jets with
  Spinning Black Holes}.
\newblock {\em Science}, 339(6115):49, January 2013.

\bibitem{Meyer94}
F.~{Meyer} and E.~{Meyer-Hofmeister}.
\newblock {Accretion disk evaporation by a coronal siphon flow.}
\newblock {\em \aap}, 288:175--182, August 1994.

\bibitem{Middleton18}
M.~J. {Middleton}, P.~C. {Fragile}, M.~{Bachetti}, M.~{Brightman}, Y.~F.
  {Jiang}, W.~C.~G. {Ho}, T.~P. {Roberts}, A.~R. {Ingram}, T.~{Dauser},
  C.~{Pinto}, D.~J. {Walton}, F.~{Fuerst}, A.~C. {Fabian}, and N.~{Gehrels}.
\newblock {Lense-Thirring precession in ULXs as a possible means to constrain
  the neutron star equation of state}.
\newblock {\em \mnras}, 475(1):154--166, March 2018.

\bibitem{Middleton16}
Matthew~J. {Middleton}, Michael~L. {Parker}, Christopher~S. {Reynolds},
  Andrew~C. {Fabian}, and Anne~M. {Lohfink}.
\newblock {The view of AGN-host alignment via reflection spectroscopy}.
\newblock {\em \mnras}, 457(2):1568--1576, April 2016.

\bibitem{Miller02}
J.~M. {Miller}, A.~C. {Fabian}, J.~J.~M. {in't Zand}, C.~S. {Reynolds},
  R.~{Wijnands}, M.~A. {Nowak}, and W.~H.~G. {Lewin}.
\newblock {A Relativistic Fe K{\ensuremath{\alpha}} Emission Line in the
  Intermediate-Luminosity BeppoSAX Spectrum of the Galactic Microquasar V4641
  Sgr}.
\newblock {\em \apjl}, 577(1):L15--L18, September 2002.

\bibitem{Miller09}
J.~M. {Miller}, C.~S. {Reynolds}, A.~C. {Fabian}, G.~{Miniutti}, and L.~C.
  {Gallo}.
\newblock {Stellar-Mass Black Hole Spin Constraints from Disk Reflection and
  Continuum Modeling}.
\newblock {\em \apj}, 697(1):900--912, May 2009.

\bibitem{Miller-Jones19}
James C.~A. {Miller-Jones}, Alexandra~J. {Tetarenko}, Gregory~R. {Sivakoff},
  Matthew~J. {Middleton}, Diego {Altamirano}, Gemma~E. {Anderson}, Tomaso~M.
  {Belloni}, Rob~P. {Fender}, Peter~G. {Jonker}, Elmar~G. {K{\"o}rding},
  Hans~A. {Krimm}, Dipankar {Maitra}, Sera {Markoff}, Simone {Migliari},
  Kunal~P. {Mooley}, Michael~P. {Rupen}, David~M. {Russell}, Thomas~D.
  {Russell}, Craig~L. {Sarazin}, Roberto {Soria}, and Valeriu {Tudose}.
\newblock {A rapidly changing jet orientation in the stellar-mass black-hole
  system V404 Cygni}.
\newblock {\em \nat}, 569(7756):374--377, April 2019.

\bibitem{Mishra19}
Bhupendra {Mishra}, W{\l}odek {Klu{\'z}niak}, and P.~Chris {Fragile}.
\newblock {Breathing Oscillations in a Global Simulation of a Thin Accretion
  Disk}.
\newblock {\em \mnras}, 483(4):4811--4819, March 2019.

\bibitem{Teixeira14}
Danilo {Morales Teixeira}, P.~Chris {Fragile}, Viacheslav~V. {Zhuravlev}, and
  Pavel~B. {Ivanov}.
\newblock {Conservative GRMHD Simulations of Moderately Thin, Tilted Accretion
  Disks}.
\newblock {\em \apj}, 796(2):103, December 2014.

\bibitem{Motta15}
S.~E. {Motta}, P.~{Casella}, M.~{Henze}, T.~{Mu{\~n}oz-Darias}, A.~{Sanna},
  R.~{Fender}, and T.~{Belloni}.
\newblock {Geometrical constraints on the origin of timing signals from black
  holes}.
\newblock {\em \mnras}, 447(2):2059--2072, February 2015.

\bibitem{Motta18}
S.~E. {Motta}, A.~{Franchini}, G.~{Lodato}, and G.~{Mastroserio}.
\newblock {On the different flavours of Lense-Thirring precession around
  accreting stellar mass black holes}.
\newblock {\em \mnras}, 473(1):431--439, January 2018.

\bibitem{Musoke23}
G.~{Musoke}, M.~{Liska}, O.~{Porth}, Michiel {van der Klis}, and Adam {Ingram}.
\newblock {Disc tearing leads to low and high frequency quasi-periodic
  oscillations in a GRMHD simulation of a thin accretion disc}.
\newblock {\em \mnras}, 518(2):1656--1671, January 2023.

\bibitem{Narayan03}
Ramesh {Narayan}, Igor~V. {Igumenshchev}, and Marek~A. {Abramowicz}.
\newblock {Magnetically Arrested Disk: an Energetically Efficient Accretion
  Flow}.
\newblock {\em \pasj}, 55:L69--L72, December 2003.

\bibitem{Narayan12}
Ramesh {Narayan}, Aleksander {S{\"A} dowski}, Robert~F. {Penna}, and Akshay~K.
  {Kulkarni}.
\newblock {GRMHD simulations of magnetized advection-dominated accretion on a
  non-spinning black hole: role of outflows}.
\newblock {\em \mnras}, 426(4):3241--3259, November 2012.

\bibitem{Nealon15}
R.~{Nealon}, D.~J. {Price}, and C.~J. {Nixon}.
\newblock {On the Bardeen-Petterson effect in black hole accretion discs}.
\newblock {\em \mnras}, 448:1526--1540, April 2015.

\bibitem{Nelson00}
Richard~P. {Nelson} and John C.~B. {Papaloizou}.
\newblock {Hydrodynamic simulations of the Bardeen-Petterson effect}.
\newblock {\em \mnras}, 315(3):570--586, July 2000.

\bibitem{Nixon12b}
Chris {Nixon}, Andrew {King}, Daniel {Price}, and Juhan {Frank}.
\newblock {Tearing up the Disk: How Black Holes Accrete}.
\newblock {\em \apjl}, 757(2):L24, October 2012.

\bibitem{Orosz97}
Jerome~A. {Orosz} and Charles~D. {Bailyn}.
\newblock {Optical Observations of GRO J1655-40 in Quiescence. I. A Precise
  Mass for the Black Hole Primary}.
\newblock {\em \apj}, 477(2):876--896, March 1997.

\bibitem{Orosz02}
Jerome~A. {Orosz}, Paul~J. {Groot}, Michiel {van der Klis}, Jeffrey~E.
  {McClintock}, Michael~R. {Garcia}, Ping {Zhao}, Raj~K. {Jain}, Charles~D.
  {Bailyn}, and Ronald~A. {Remillard}.
\newblock {Dynamical Evidence for a Black Hole in the Microquasar XTE
  J1550-564}.
\newblock {\em \apj}, 568(2):845--861, April 2002.

\bibitem{Papaloizou95}
J.~C.~B. {Papaloizou} and D.~N.~C. {Lin}.
\newblock {On the Dynamics of Warped Accretion Disks}.
\newblock {\em \apj}, 438:841, January 1995.

\bibitem{Papaloizou83}
J.~C.~B. {Papaloizou} and J.~E. {Pringle}.
\newblock {The time-dependence of non-planar accretion discs}.
\newblock {\em \mnras}, 202:1181--1194, March 1983.

\bibitem{Pasham19}
Dheeraj~R. {Pasham}, Ronald~A. {Remillard}, P.~Chris {Fragile}, Alessia
  {Franchini}, Nicholas~C. {Stone}, Giuseppe {Lodato}, Jeroen {Homan}, Deepto
  {Chakrabarty}, Frederick~K. {Baganoff}, James~F. {Steiner}, Eric~R.
  {Coughlin}, and Nishanth~R. {Pasham}.
\newblock {A loud quasi-periodic oscillation after a star is disrupted by a
  massive black hole}.
\newblock {\em Science}, 363(6426):531--534, February 2019.

\bibitem{Pasham24}
Dheeraj~R. {Pasham}, Michal {Zajacek}, C.~J. {Nixon}, Eric~R. {Coughlin},
  Marzena {Sniegowska}, Agnieszka {Janiuk}, Bozena {Czerny}, Thomas {Wevers},
  Muryel {Guolo}, Yukta {Ajay}, and Michael {Loewenstein}.
\newblock {Lense-Thirring Precession after a Supermassive Black Hole Disrupts a
  Star}.
\newblock {\em arXiv e-prints}, page arXiv:2402.09689, February 2024.

\bibitem{Poutanen22}
Juri {Poutanen}, Alexandra {Veledina}, Andrei~V. {Berdyugin}, Svetlana~V.
  {Berdyugina}, Helen {Jermak}, Peter~G. {Jonker}, Jari J.~E. {Kajava}, Ilia~A.
  {Kosenkov}, Vadim {Kravtsov}, Vilppu {Piirola}, Manisha {Shrestha}, Manuel~A.
  {Perez Torres}, and Sergey~S. {Tsygankov}.
\newblock {Black hole spin{\textendash}orbit misalignment in the x-ray binary
  MAXI J1820+070}.
\newblock {\em Science}, 375(6583):874--876, February 2022.

\bibitem{Pringle96}
J.~E. {Pringle}.
\newblock {Self-induced warping of accretion discs}.
\newblock {\em \mnras}, 281(1):357--361, July 1996.

\bibitem{Qian07}
Lei Qian, B.~F. Liu, and Xue-Bing Wu.
\newblock Disk evaporation-fed corona: Structure and evaporation features with
  magnetic field.
\newblock {\em The Astrophysical Journal}, 668(2):1145, 2007.

\bibitem{Rees78}
M.~J. {Rees}.
\newblock {Relativistic jets and beams in radio galaxies}.
\newblock {\em \nat}, 275(5680):516--517, October 1978.

\bibitem{Reis10}
R.~C. {Reis}, A.~C. {Fabian}, and J.~M. {Miller}.
\newblock {Black hole accretion discs in the canonical low-hard state}.
\newblock {\em \mnras}, 402(2):836--854, February 2010.

\bibitem{Ressler18}
S.~M. {Ressler}, E.~{Quataert}, and J.~M. {Stone}.
\newblock {Hydrodynamic simulations of the inner accretion flow of Sagittarius
  A* fuelled by stellar winds}.
\newblock {\em \mnras}, 478(3):3544--3563, August 2018.

\bibitem{Ressler23}
S.~M. {Ressler}, C.~J. {White}, and E.~{Quataert}.
\newblock {Wind-fed GRMHD simulations of Sagittarius A*: tilt and alignment of
  jets and accretion discs, electron thermodynamics, and multiscale modelling
  of the rotation measure}.
\newblock {\em \mnras}, 521(3):4277--4298, May 2023.

\bibitem{Reynolds09}
Christopher~S. {Reynolds} and M.~Coleman {Miller}.
\newblock {The Time Variability of Geometrically Thin Black Hole Accretion
  Disks. I. The Search for Modes in Simulated Disks}.
\newblock {\em \apj}, 692(1):869--886, February 2009.

\bibitem{Scheuer96}
P.~A.~G. {Scheuer} and R.~{Feiler}.
\newblock {The realignment of a black hole misaligned with its accretion disc}.
\newblock {\em \mnras}, 282:291, September 1996.

\bibitem{Schmitt02}
H.~R. {Schmitt}, J.~E. {Pringle}, C.~J. {Clarke}, and A.~L. {Kinney}.
\newblock {The Orientation of Jets Relative to Dust Disks in Radio Galaxies}.
\newblock {\em \apj}, 575(1):150--155, August 2002.

\bibitem{Shakura73}
N.~I. {Shakura} and R.~A. {Sunyaev}.
\newblock {Black holes in binary systems. Observational appearance.}
\newblock {\em \aap}, 24:337--355, January 1973.

\bibitem{Sironi24}
Lorenzo {Sironi} and Aaron {Tran}.
\newblock {Electron Heating in the Trans-Relativistic Perpendicular Shocks of
  Tilted Accretion Flows}.
\newblock {\em arXiv e-prints}, page arXiv:2402.13317, February 2024.

\bibitem{Sorathia13}
Kareem~A. {Sorathia}, Julian~H. {Krolik}, and John~F. {Hawley}.
\newblock {Magnetohydrodynamic Simulation of a Disk Subjected to Lense-Thirring
  Precession}.
\newblock {\em \apj}, 777(1):21, November 2013.

\bibitem{Tchekhovskoy11}
Alexander {Tchekhovskoy}, Ramesh {Narayan}, and Jonathan~C. {McKinney}.
\newblock {Efficient generation of jets from magnetically arrested accretion on
  a rapidly spinning black hole}.
\newblock {\em \mnras}, 418(1):L79--L83, November 2011.

\bibitem{White19}
Christopher~J. {White}, Eliot {Quataert}, and Omer {Blaes}.
\newblock {Tilted Disks around Black Holes: A Numerical Parameter Survey for
  Spin and Inclination Angle}.
\newblock {\em \apj}, 878(1):51, June 2019.

\end{thebibliography}
\bibliographystyle{plain}


\end{document}